\documentclass[fleqn,usenatbib]{mnras}

\usepackage[version=4]{mhchem}
\usepackage[T1]{fontenc}

\DeclareRobustCommand{\VAN}[3]{#2}
\let\VANthebibliography\thebibliography
\def\thebibliography{\DeclareRobustCommand{\VAN}[3]{##3}\VANthebibliography}

\usepackage{graphicx}	% Including figure files
\usepackage{amsmath}	% Advanced maths commands
\usepackage{amssymb}	% Extra maths symbols
\usepackage{comment}
\usepackage{newtxtext,newtxmath}

\title[Silicate vapor in sub-Neptunes]{The importance of silicate vapor in determining the structure, radii, and envelope mass fractions of sub-Neptunes}

\author[W. Misener and H.E. Schlichting]{
William Misener$^{1}$
and Hilke E. Schlichting$^{1}$
\\
$^{1}$Department of Earth, Planetary, and Space Sciences, The University of California, Los Angeles, 595 Charles E. Young Drive East, Los Angeles, CA 90095, USA\\}

\date{Accepted XXX. Received YYY; in original form ZZZ}

\pubyear{2022}

\begin{document}
\label{firstpage}
\pagerange{\pageref{firstpage}--\pageref{lastpage}}
\maketitle

% Abstract of the paper
\begin{abstract}
Substantial silicate vapor is expected to be in chemical equilibrium at temperature conditions typical of the silicate-atmosphere interface of sub-Neptune planets, which can exceed 5000~K. Previous models of the atmospheric structure and evolution of these exoplanets, which have been used to constrain their atmospheric mass fractions, have neglected this compositional coupling. In this work, we show that silicate vapor in a hydrogen-dominated atmosphere acts as a condensable species, decreasing in abundance with altitude. The resultant mean molecular weight gradient inhibits convection at temperatures above $\sim 4000$~K, inducing a near-surface radiative layer. This radiative layer decreases the planet's total radius compared to a planet with the same base temperature and a convective, pure H/He atmosphere. Therefore, we expect silicate vapor to have major effects on the inferred envelope mass fraction and thermal evolution of sub-Neptune planets. We demonstrate that differences in radii, and hence in inferred atmospheric masses, are largest for planets which have larger masses, equilibrium temperatures, and atmospheric mass fractions. The effects are largest for younger planets, but differences can persist on gigayear time-scales for some sub-Neptunes. For a $10 M_\oplus$ planet with $T_\mathrm{eq}=1000$~K and an age of $\sim 300$~Myr, an observed radius consistent with an atmospheric mass fraction of 10\% when accounting for silicate vapor would be misinterpreted as indicating an atmospheric mass fraction of 2\% if a H/He-only atmosphere were assumed. The presence of silicate vapor in the atmosphere is also expected to have important implications for the accretion and loss of primordial hydrogen atmospheres. %[248 words, max 250]
\end{abstract}

% Select between one and six entries from the list of approved keywords.
% Don't make up new ones.

%%%%%%%%%%%%%%%%%%%%%%%%%%%%%%%%%%%%%%%%%%%%%%%%%%

%%%%%%%%%%%%%%%%% BODY OF PAPER %%%%%%%%%%%%%%%%%%

\section{Introduction}
Exoplanet survey missions have revealed that the most common planets orbiting in less than 100 days around their host stars have radii in between those of Earth and Neptune \citep[e.g.][]{F13}. Further refinements in these planets' properties have revealed that this population has a bimodal radius and density distribution, dividing the planets into hydrogen gas-rich sub-Neptunes and gas-poor super-Earths \citep{WM14,FP17}.

The ubiquity of sub-Neptunes in the Galaxy has prompted numerous works investigating the physical and chemical mechanisms that determine their radii as we observe them today. A clear outcome is that the radius of a planet with a thick hydrogen envelope is not constant with age, but shrinks with time as the planet radiates energy and cools \citep[e.g.][]{LF14}. Furthermore, the atmospheric mass of a planet may not be constant in time. A common hypothesis to explain the radius dichotomy is that super-Earths and sub-Neptunes began as a single population of silicate-dominated cores which accreted a few percent of their mass in nebular hydrogen and helium. Subsequent processes, such as photo-evaporation \citep[e.g.][]{OJ12} and core-powered mass loss \citep{GSS16}, then drove atmospheric mass loss, completely stripping the eventual super-Earths of their primordial gas. Sub-Neptunes were able to mostly resist this stripping and remain today as silicate-dominated cores overlain by a thick hydrogen atmosphere. Both photo-evaporation and core-powered mass loss can explain trends in the observed radius distribution \citep[e.g.][]{OW17, GS19}, with no clear consensus on which is the dominant mechanism \citep{LS20, RogersGupta2021}.

However, attempts at modeling these planets' evolution reveal that the problem of inferring internal  composition from the bulk densities of even the most well-characterized small planets is highly degenerate. Many sub-Neptunes can be modeled as containing solely silicates, metals, and hydrogen, but their observed properties can also accommodate significant fractions of volatiles, such as water \citep[e.g.][]{RS10,D17}, which have densities intermediate to hydrogen gas and silicate. Whether one expects such volatile fractions depends in part on the formation location of these planets' cores in the protoplanetary disk, which is not currently clear. One constraint on the volatile fractions comes from mass loss models, as lower density cores would more easily lose their hydrogen envelopes. Atmospheric mass loss models are most consistent with observations if cores are Earth-like in composition, with less than 20 percent of their mass in icy material \citep{GS19,RO21}. The densities of ultra-short period planets are also consistent with predominantly rocky composition \citep{Dai19}. Nevertheless, the bulk composition of sub-Neptunes remains uncertain, as formation and interior models can also reproduce the sizes of sub-Neptune planets by modeling them as hydrogen-poor and water-rich \citep[e.g.][]{Zeng2019, Mousis2020}.

Extensive work has been done on inferring the internal compositions of well-characterized exoplanets, that is, those with precisely measured masses and radii, such as those orbiting TRAPPIST-1 \citep[e.g.][]{U18} and K2-18 \citep{M20}. Such models tend to assume a layered structure for the interiors of these planets, with a pure metal core at the center, a rocky silicate mantle, a potential high-pressure ice layer, and finally a gaseous hydrogen layer \citep[e.g.][]{LF14, D17}. In reality, however, these layers may not be so discrete. For example, recent \textit{Juno} measurements have provided evidence that Jupiter's assumed high-density core may be dilute rather than segregated from its surroundings \citep{W17Juno}. While the conditions at the center of Jupiter are in a different pressure regime than sub-Neptunes, recent work has shown that under the interior pressures typical of sub-Neptunes, water and rock may be significantly miscible \citep{VazanSari22}, calling in to question the discrete layers typically assumed.

Similarly to the deep interior, the outer hydrogen envelope is not isolated from its underlying layers either. Instead, there may be significant interaction at the surface of the silicate mantle. Experiments on hydrogen solubility in silicate melt indicate that significant hydrogen may in-gas into the mantle \citep{H12}. This in-gassing could have important effects on the planets' long-term evolution, potentially buffering atmospheric loss via gradual outgassing \citep[e.g.][]{CS18} and leading to the migration of light elements to the metal core \citep{SY21}. Similarly, the chemical coupling of the silicate mantle and atmosphere could lead to out-gassing of interior volatiles. These volatiles could have significant effects on the atmospheric structure in a number of ways. First, their absorption characteristics can affect the atmosphere's overall radiative properties, which can change how the atmosphere cools with time as well as modify the observable spectral features. Such observable features are also affected by the volatile species' large mean molecular weights compared to hydrogen, which decrease the atmosphere's scale height. Additionally, if the outgassed species can condense somewhere in the atmosphere, this could modify the atmosphere's structure due to the release of latent heat \citep[e.g.][]{L21}.

Since gas can hold more of the condensable species at higher temperature, abundant condensable species can set up a mean molecular weight gradient within the atmosphere. If the gradient becomes too steep, convection can be inhibited, failing to satisfy the Ledoux criterion \citep{L47}. This can lead to development of a radiative region deep in the atmosphere. This inhibition of convection has been studied in the context of the Solar System gas and ice giants  \citep{G95,L17,VHG18,Markham21}. The typical condensables considered in these previous studies are water and methane. Such a radiative region may also be caused by entropy gradients that arise during accretion \citep{CHV18}.

Recent work has found that at the time of their formation, the surfaces of the silicate cores of sub-Neptunes are expected to be very hot, over $10^4$ K \citep[e.g.][]{GSS16}. At chemical equilibrium at such large temperatures, substantial silicate vapor is stable in the gas phase \citep{FS12,VF13,BS21,SY21}. This silicate vapor then condenses as the temperature of the atmosphere drops with increasing altitude, with condensation being mostly complete once temperatures are below 2000~K. Therefore, for young sub-Neptunes, silicate vapor acts as a condensable, with a large reservoir available at the base of the atmosphere. \citet{BS21} suggested that the presence of such silicates could significantly change the overall mean molecular weight of the atmosphere, but that this effect may be counteracted if the vapor is kept concentrated near the surface. Some works, focused on accretion processes, have also considered the effect of this silicate vapor, finding that molecular weight gradients form as pebbles evaporate during infall \citep{BrouwersOrmel2020}. Subsequent structural modeling concluded that convection could be inhibited by these molecular weight gradients \citep{OrmelVazan2021}, but these works consider only the disk-phase of accretion, not the long-term effects on a planet's thermal evolution and the resulting changes in the corresponding mass-radius relation.

In this work, we model the atmospheric structure and evolution of sub-Neptune planets, accounting for the presence of silicate vapor in equilibrium with an underlying magma ocean. We demonstrate that the condensation of silicate vapor has a major effect on the overall radii of sub-Neptunes, especially at young ages. The silicate vapor is sufficiently abundant at the base of the atmosphere to form a mean molecular weight gradient, independent of the formation mechanism of the planet. Such a gradient inhibits convection and forms a radiative layer at the base of the atmosphere. The size of the radiative layer depends sensitively on the opacity, but we show that using common values for high density hydrogen leads to a very steep temperature gradient. This has the effect of significantly shrinking the envelope width compared to a fully adiabatic atmosphere with the same surface temperature. We outline our methods in Section~\ref{sec:methods}, present our results regarding the atmospheric structure and evolution of these planets and discuss their implications for inferred atmospheric masses in Section~\ref{sec:results}, analyze key assumptions and the prospects for future work in Section~\ref{sec:discussion}, and summarize our conclusions in Section~\ref{sec:conc}.

\section{Model and Approach}\label{sec:methods}
In this section, we describe our model. We detail the key parameters that govern sub-Neptune atmospheres, including our model of the core, the atmosphere, and their evolution in time. Many of the basic model parameters follow those used in \citet{MS21}, with modifications to include silicate vapor and to ignore any mass-loss.

\subsection{Interior model}\label{sec:core}
Motivated by comparisons of atmospheric loss models to observed demographics \citep{GS19,RO21} and the measured densities of ultra-short period planets \citep{Dai19}, in this work we model sub-Neptunes as consisting of a silicate magma ocean core in contact with a hydrogen-rich atmosphere. We assume these cores to have densities consistent with the silicate-metal composition of Earth, accounting for compression. However, we note that observations do not preclude potentially large variations in bulk water content between individual planets and systems \citep[e.g.][]{RS10,D17,Zeng2019,Mousis2020}. We discuss the potential effects of volatile species such as water in Section \ref{sec:cond}. The core radius, $R_\mathrm{c}$, is determined by the core mass, $M_\mathrm{c}$, via the mass-radius relation $R_\mathrm{c}/R_{\earth} = (M_\mathrm{c}/M_{\earth})^{1/\beta}$, where $R_{\earth}$ and $M_{\earth}$ are the radius and mass of Earth, respectively. A power law scaling of $\beta \simeq 4$ has been shown to be in good agreement with internal structure models of super-Earths with Earth-like compositions \citep[e.g.][]{V06, S07}. Since we focus on sub-Neptunes, which are thought to have atmospheric masses on the order of a few percent of the planets' total masses, we ignore the atmosphere's contribution to the gravity field.

We take the energy available for cooling in the silicate core as its thermal energy,
\begin{equation}
    E_\mathrm{c} = C_\mathrm{c} T_\mathrm{c} \simeq \frac{1}{\gamma_\mathrm{c}-1} N k_\mathrm{B} T_\mathrm{c}.
	\label{eq:core_E}
\end{equation}
Here, $C_\mathrm{c}$ is the heat capacity of the core and $T_\mathrm{c}$ is the temperature, which we take as coupled to the temperature at the base of the atmosphere. This energy formulation assumes the silicate interior is isothermal, though incorporating the true thermal gradient within the core would not significantly change its overall heat capacity. In the second equality, we model the heat capacity in the form of an ideal gas, where $k_\mathrm{B}$ is the Boltzmann constant. We assume throughout this work that the effective adiabatic index of the core $\gamma_\mathrm{c}=4/3$, following the Dulong-Petit law. As liquid silicate can have a higher heat capacity than the solid form \citep[e.g.][]{SSD17}, this value for the adiabatic index represents an upper limit. The number of molecules in the core $N=M_\mathrm{c}/\mu_\mathrm{c}$, where $\mu_\mathrm{c}$ is the mean molecular weight of the core. We assume $\mu_\mathrm{c}=60$ amu, similar to that of Earth.

\subsection{Atmospheric Structure}\label{sec:atm_struc}
We model the atmospheric structure as containing an outer isothermal radiative region, which transitions to a convective region at the radiative-convective boundary, $R_\mathrm{rcb}$ \citep[e.g.][]{LC15, GSS16, MS21}. This outer region, and therefore the radiative-convective boundary, is in thermal equilibrium with the incident stellar radiative flux and therefore has a temperature, $T_\mathrm{eq}$, that scales with the planet's semi-major axis, $a$:

\begin{equation}\label{eq:tempa}
    T_\mathrm{eq} = \bigg(\frac{L_*}{16 \pi \sigma a^2}\bigg)^{1/4} = 279\ \mathrm{K} \bigg(\frac{a}{1~\mathrm{au}}\bigg)^{-1/2}\mathrm{,}
\end{equation}
where $L_*$ is the luminosity of the planet's host star, $\sigma$ is the Stefan-Boltzmann constant, and the right-most expression is evaluated assuming a Sun-like stellar host, using $L_*=L_\odot$. This result matches the classic \citet{H81} profile. The upper isothermal region has negligible mass compared to the regions below \citep{MS21}, and so we assume the entire mass of the atmosphere is contained inside the radiative-convective boundary.

To construct the atmospheric profile, we increase the pressure from the radiative-convective boundary in small pressure steps, $\Delta P$. After each of these steps, we calculate the new pressure, $P_\mathrm{new}=P + \Delta P$, the new radius, $R_\mathrm{new} = R + \Delta P/(\partial P/\partial R)$, and the new temperature, $T_\mathrm{new} = T + \Delta P (\partial T/\partial P)$. The change in radius over the small pressure step is found assuming hydrostatic equilibrium:
\begin{equation}\label{eq:dPdr}
    \frac{\partial P}{\partial R} = -\frac{G M_\mathrm{c}}{R^2} \frac{\mu P}{k_\mathrm{B} T},
\end{equation}
where $G$ is the gravitational constant and $\mu$ is the local mean molecular weight. The temperature gradient, as well as the radial gradient through its dependence on the mean molecular weight, both depend on the abundance of silicate vapor in the atmosphere, which we detail in the next section.

In order to fully determine an atmosphere, we must specify the planet's core mass and equilibrium temperature, the mass of the atmosphere, $M_\mathrm{atm}$, and the planet's total available energy for cooling, $E$. We then solve for the atmospheric profile, including the radiative-convective boundary radius, that satisfies these boundary conditions. The atmospheric mass is integrated from its density profile:

\begin{equation}\label{eq:mass_int}
    M_\mathrm{atm} = \int_{R_\mathrm{c}}^{R_\mathrm{rcb}} 4 \pi R^2 \rho(R) dR,
\end{equation}
where the density $\rho=\mu P/(k_\mathrm{B} T)$.
Similarly, the total energy available for cooling is the sum of the available energy of the core and atmosphere: $E=E_\mathrm{c}+E_\mathrm{atm}$. Since we model the core as incompressible, we neglect its gravitational potential energy, and $E_\mathrm{c}$ is given by the thermal energy presented in equation~(\ref{eq:core_E}). Meanwhile, the atmosphere's total energy is
\begin{equation}\label{eq:energy_int}
    E_\mathrm{atm} = \int_{M_\mathrm{atm}} e dm \simeq \int_{R_\mathrm{c}}^{R_\mathrm{rcb}} 4 \pi R^2 e(R) \rho(R) dR,
\end{equation}
where $e$ is the atmospheric specific energy, the sum of the (negative) gravitational and (positive) thermal energy:
\begin{equation}\label{eq:energy_specific}
    e(R) = -\frac{G M_\mathrm{c}}{R} + \frac{1}{\gamma-1} \frac{k_\mathrm{B} T}{\mu}.
\end{equation}

\subsubsection{Effects of silicate vapor}\label{sec:silicate_struc}
Condensable minor species can affect atmospheric structure in two main ways. First, the release of the latent heat of condensation as the atmosphere convects changes the energy balance and therefore the profile of the atmosphere. Second, as hotter gas can hold more condensate, a mean molecular weight gradient as a function of temperature can develop. In hydrogen-dominated atmospheres, since condensates will have larger molecular weights than the hydrogen gas, this molecular gradient will increase with depth, which acts against convective instability. \citet{L17} examined the effects of the condensables water and methane on the structure of Uranus and Neptune. They found that at the typical temperatures of these planets, water could fully inhibit convection in their atmospheres, forming a radiative region with a steep temperature gradient.

The conditions in young, close-in sub-Neptunes typical of the known exoplanet population are different than the Solar System ice giants. Their equilibrium temperatures are usually hotter than the boiling temperature of water; in fact, they can approach $\sim 2000$~K, a typical sublimation temperature for rock, and their interiors can easily exceed this, especially when young. Here we apply the equations derived in \citet{L17}, using silicates as the condensate rather than a more volatile species.

We assume that the atmosphere is in thermal and chemical equilibrium with the underlying silicate interior. Thermal equilibrium implies that the temperature at the base of the atmosphere is the same as that of the silicate core, $T_\mathrm{c}=T(R_\mathrm{c})$, and that for the atmosphere to cool, the core must also cool. This coupling is the basis of core-powered mass-loss \citep[e.g.][]{GSS16, MS21}, though we do not consider mass-loss in this work. Meanwhile, chemical equilibrium implies that the atmosphere is saturated in silicate vapor. For simplicity, we assume all silicate vapor is in the form of SiO. More detailed chemical equilibrium calculations, such as those using the \textsc{magma} code \citep{FC87, SF04}, find that SiO predominates at equilibrium over much of the temperature range we consider here. However, in equilibrium with pure silicate, there can also be substantial amounts of \ce{O2}, O, and \ce{SiO2}. In addition, we do not consider any other elements besides silicon and oxygen. At temperatures below 2500~K, gas in equilibrium with bulk silicate Earth composition includes Na and Zn gas in addition to silicate vapor \citep[e.g.][]{VF13}. However, full chemical equilibrium calculations are beyond the scope of this study, so we do not include the effects of the less refractory species. 

For the vapor pressure of silicate gas, $P_\mathrm{svp}$, we take the expression of \citet{FS12} and \citet{VF13}, based on the \textsc{magma} code:
\begin{equation}\label{eq:Psvp}
    \log_{10} \bigg(\frac{P_\mathrm{svp}}{\mathrm{bar}}\bigg) = 8.203 - \frac{25898.9}{T/\mathrm{K}}.
\end{equation}
Due to our neglect of less refractory species, this approximation underestimates the true vapor pressure of bulk silicate Earth species at temperatures below 2500~K, with errors becoming more substantial at lower temperatures. However, since the overall vapor pressure is low at low temperatures, this approximation does not have a substantial effect on our overall findings.
This method also overestimates the true bulk silicate Earth vapor pressure by a factor of a few at temperatures above 2500~K \citep{VF13}. No matter the exact approximation, the key feature of the vapor pressure equation is that it is a strongly increasing function of temperature.

In the convective region, silicate vapor modifies the profile due to the release of latent heat. Its condensation causes the overall atmospheric profile to follow the so-called wet adiabat, well-studied in the context of Earth's atmosphere. The general temperature gradient can be expressed as:
\begin{equation}\label{eq:wetadiabat}
    \frac{\partial \ln T}{\partial \ln P} = \frac{k_\mathrm{B}}{\mu} \cfrac{1+\cfrac{P_\mathrm{svp}}{P_\mathrm{H}} \cfrac{\partial \ln P_\mathrm{svp}} {\partial T}} {c_\mathrm{p} +\cfrac{P_\mathrm{svp}} {P_\mathrm{H}} \cfrac{k_\mathrm{B}}{\mu} T^2 \bigg(\cfrac{\partial \ln P_\mathrm{svp}} {\partial T}\bigg)^2},
\end{equation}
where $P_\mathrm{H} = P - P_\mathrm{svp}$ is the partial pressure of hydrogen/helium \citep[e.g.][]{L17}. Following from this definition, the local mean molecular weight, $\mu$, is:
\begin{equation}\label{eq:mmw}
    \mu = \frac{\mu_\mathrm{H} P_\mathrm{H} + \mu_\mathrm{sv} P_\mathrm{svp}}{P},
\end{equation}
where we take $\mu_\mathrm{H} = 2.4$~amu, typical for a nebular hydrogen/helium mixture. The mean molecular weight of pure SiO vapor is 44~amu. However, detailed chemical equilibrium simulations, e.g., those performed by the \textsc{magma} code, find that the mean molecular weight of vapor at equilibrium with a $\sim 6000$~K magma ocean is slightly lower, $\mu_\mathrm{sv} \sim 36$~amu, due to the presence of atomic Si and O. In order to capture the effects of these lower weight components, we use this latter value as the condensable mean molecular weight \citep[e.g.][]{BS21}. However, since SiO still dominates the composition, throughout this work we use its thermodynamic properties for all other calculations. We find the heat capacity of the mixture, $c_\mathrm{p}$, to be
\begin{equation}\label{eq:heatcap}
    c_\mathrm{p} = \frac{k_\mathrm{B}}{\mu} \frac{\gamma}{\gamma-1},
\end{equation}
where $\gamma$ is the adiabatic index. We assume ideal gas behavior, which for the diatomic molecules SiO and \ce{H2} yields $\gamma=7/5$. It is easy to see that equation~(\ref{eq:wetadiabat}) reduces to the dry adiabatic case, $\partial \ln T /\partial \ln P \simeq (\gamma-1)/\gamma$, when $P_\mathrm{svp} \ll P_\mathrm{H}$.

If the condensable vapor pressure increases enough, it will begin to have an effect on the overall mean molecular weight of the atmosphere. In an atmosphere primarily composed of a light species, such as hydrogen, this acts to increase the mean molecular weight as the temperature increases. This increases the stability of the gas against convection, as a lifted gas parcel will have a larger mean molecular weight than its surroundings. The balance between the temperature gradient and the molecular weight gradient depends on the mass mixing ratio, $q=\mu_\mathrm{sv} P_\mathrm{svp}/(\mu_\mathrm{H} P_\mathrm{H})$. Once the temperature increases such that the gas reaches a critical mass mixing ratio, $q \geq q_\mathrm{crit}$, convection is no longer possible. Convection is inhibited no matter how super-adiabatic the temperature gradient becomes: an increase in the temperature gradient merely increases the molecular weight gradient. Therefore, an inner radiative region forms \citep{G95}. Such a region is also stable to double-diffusive convection \citep{L17}.

This critical mixing ratio, $q_\mathrm{crit}$, depends on the weights of the constituent species and on the vapor pressure gradient \citep{G95, L17}

\begin{equation}\label{eq:qcrit}
    q_\mathrm{crit} = \cfrac{1}{\bigg(1-\cfrac{\mu_\mathrm{H}}{\mu_\mathrm{sv}}\bigg) \cfrac{\partial \ln P_\mathrm{svp}}{\partial \ln T}}.
\end{equation}
The critical mixing ratio increases slowly with temperature, and is typically on the order of 10 weight percent in sub-Neptune atmospheres. This value is of the same order of magnitude as the value for other volatiles in Solar System gas giants \citep{L17}.

Below the location at which this mixing ratio is achieved, the atmosphere cannot convect, and therefore must transfer energy via radiation. The energy flux that must be transported across this radiative region must in steady state be equal to the radiative flux into space. This luminosity is determined by the outer boundary conditions of the atmosphere
\begin{equation}\label{eq:luminosity}
    L = \frac{\gamma-1}{\gamma}\frac{64\pi G M_\mathrm{c} \sigma T_\mathrm{eq}^4}{3 \kappa_\mathrm{rcb} P_\mathrm{rcb}},
\end{equation}
where $\kappa_\mathrm{rcb}$ is the atmospheric Rosseland mean opacity at the radiative-convective boundary, and similarly $P_\mathrm{rcb} =P(R_\mathrm{rcb})$ \citep{GSS16}. The opacity is a function of temperature, pressure, and the atmospheric composition. At the outer boundary, typically we find $P_\mathrm{svp} \ll P$, so opacities for pure hydrogen/helium are appropriate. We use the analytic opacity approximation presented in \citet{F14}, which is based on more detailed line opacity data.

The temperature gradient necessary to transport a given energy flux, $L$, in the radiative region is:
\begin{equation}\label{eq:radtemp}
    \frac{\partial \ln T}{\partial \ln P} = \frac{3 \kappa P L}{64\pi G M_\mathrm{c} \sigma T^4}.
\end{equation}
In the interior, the opacity may be quite different than at the outer radiative-convective boundary, due not only to the typically large temperatures and pressures but also due to the highly enhanced silicate content compared to solar. Hydrogen opacity is typically taken to increase linearly with metal content, as the presence of metals enhances the production of H$^-$, the primary opacity source at high temperatures \citep[e.g.][]{LCO14}. However, this trend was not extrapolated to the high metal contents we encounter here. Therefore, the magnitude of the impact of this silicate content on the overall opacity is difficult to determine without a detailed radiative model, which is beyond the scope of this work. For simplicity, we use the same \citet{F14} relation as at the radiative-convective boundary, but acknowledge that these values are very uncertain. We discuss the effects of different opacity scalings in Sec.~\ref{sec:opacity}.

\begin{figure}
    \centering
	\includegraphics[width=0.48\textwidth]{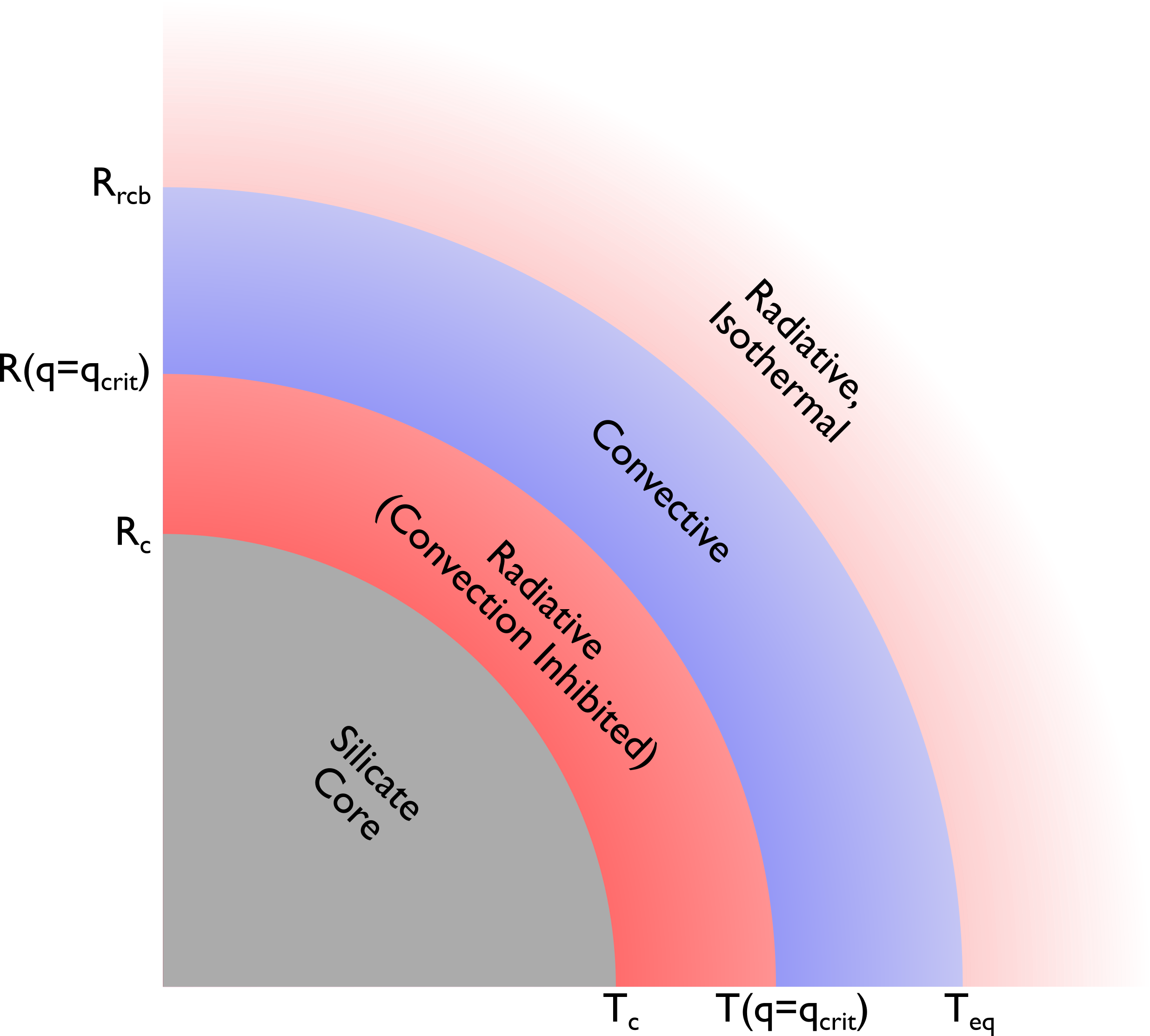}
    \caption{Illustration of the general structure of hot sub-Neptune planets. Key radii are shown on the y-axis, while key temperatures are on the x-axis. The silicate core (grey) is overlain by a hydrogen-dominated atmosphere, which is divided into radiative (red) and convective (blue) regions. An outer radiative region, isothermal at $T_\mathrm{eq}$, transitions to a convective region at the radiative-convective boundary radius, $R_\mathrm{rcb}$. If $q$ exceeds $q_\mathrm{crit}$ while $R>R_\mathrm{c}$, then convection becomes inhibited. A radiative region thus forms, extending to the base of the atmosphere.}
    \label{fig:planet_diagram}
\end{figure}

We summarize the general planet structure we find in Fig.~\ref{fig:planet_diagram}. Radiative layers are shown in red, while the convective layer is blue. The outer radiative layer is isothermal at $T_\mathrm{eq}$. It transitions at the radiative-convective boundary radius $R_\mathrm{rcb}$ to a convective region. Within the convective region, the temperature gradient follows the wet adiabat, given by Equation (\ref{eq:wetadiabat}). The mass mixing ratio $q$ increases with depth through the convective region. If $q$ exceeds $q_\mathrm{crit}$, then convection is inhibited below this radius. In this case, the atmosphere transitions to a radiative layer until it reaches the surface of the silicate core, shown in grey, at $R=R_\mathrm{c}$. Within this radiative layer, the temperature gradient follows Equation (\ref{eq:radtemp}).

Fig.~\ref{fig:planet_diagram} is not to scale, and the relative sizes of the regions depends on the assumed planet conditions. Typically, we find the energy flux and opacity to be sufficiently high that the radiative gradient is very steep: temperature increases quickly in the radiative region with pressure and radius. This steep gradient is similar to the findings of \citet{L17}. However, the inner boundary condition in the ice giants considered there is when the mixing ratio of the condensable reaches a prescribed inner value. In contrast, in sub-Neptune case we consider, the inner boundary condition is instead the surface of the silicate magma ocean, at $R=R_\mathrm{c}$, which is in our approximation entirely silicate. For some planets, this can be reached in the radiative region. However, for planets with the lowest core and atmospheric masses and highest core temperatures we consider, the hydrogen pressure in the inner regions may not be as high as the silicate vapor pressure as indicated by equation~(\ref{eq:Psvp}). In this case, we assume the atmosphere becomes composed of entirely silicate vapor, with no hydrogen gas. Such a layer is fully dry convective.

\subsection{Evolution}
We define our initial atmospheric state by its base temperature, $T_\mathrm{c}$. This value, along with the planet's core mass, equilibrium temperature, and atmospheric mass, is sufficient to define an atmospheric profile and a luminosity. A convenient method of quantifying the atmospheric evolution is through the planet's cooling timescale, $t_\mathrm{cool}$. The cooling timescale is the characteristic evolution timescale of the available energy for cooling at the current planetary luminosity, $t_\mathrm{cool}=E/L$.

To evolve these planets in time, we subtract the energy lost by the system by radiative luminosity over one timestep, $\Delta t$, defined as one-hundredth of the cooling timescale: $E_\mathrm{new} = E - L \Delta t$. We then calculate a new profile that has the new available energy, and proceed as such. 

\section{Results}\label{sec:results}
In this section we describe our main results. We begin with a general picture of the atmospheric structure we derive. We then explore the implications for the radial evolution of sub-Neptune planets. We present how our results vary as a function of planet parameters such as planet mass, atmospheric mass, equilibrium temperature, and age.

\subsection{Atmospheric structure}
\begin{figure*}
    \centering
	\includegraphics[width=\textwidth]{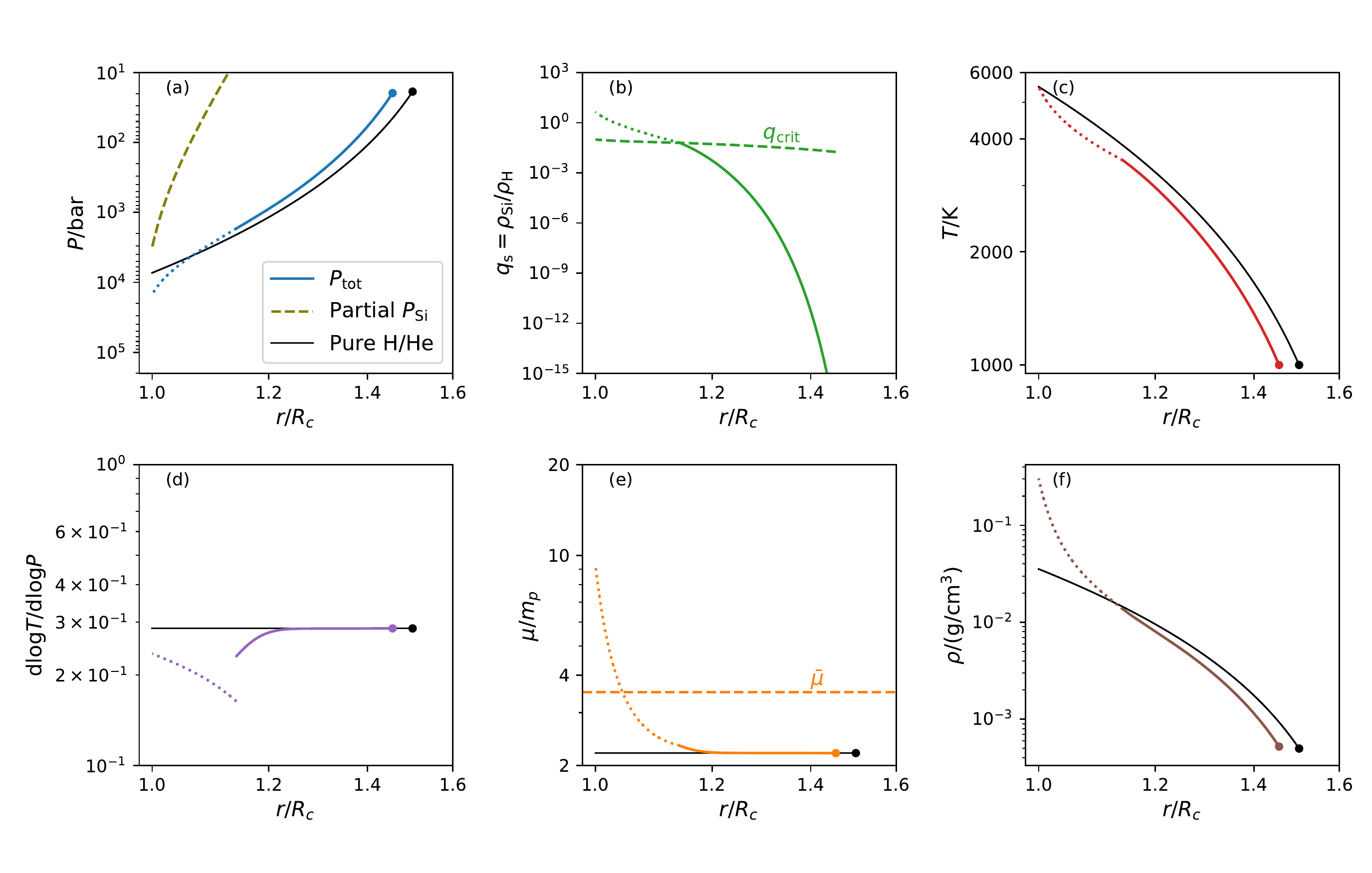}
    \caption{Atmospheric structure of a planet with a hydrogen/helium atmosphere in thermal and chemical equilibrium with a silicate core of temperature 5500~K. Coloured lines show the variation with radius $r$, of (a) total and silicate partial pressure (dashed), $P_\mathrm{tot}$ and $P_\mathrm{Si}$ respectively, (b) mass mixing ratio of silicate vapor $q_\mathrm{s}$, with the critical mixing ratio $q_\mathrm{crit}$ shown as a dashed line, (c) temperature $T$, (d) lapse rate $\mathrm{d}\log{T}/\mathrm{d}\log{P}$, (e) mean molecular weight $\mu$, with the entire atmosphere's mean molecular weight shown as a dashed line, and (f) density $\rho$. The profiles of a pure hydrogen/helium atmosphere of the same base temperature, mass, and outer temperature are shown as black lines for comparison. The outer, convective atmosphere (solid lines) has similar characteristics to the pure hydrogen case, until the critical mixing ratio is reached. Then the atmosphere becomes radiative (dotted lines), changing the temperature profile and ultimately the planet's observable radius ($R_\mathrm{rcb}$, shown by the large dot) to $1.46 R_\mathrm{c}$, compared to $R_\mathrm{rcb} = 1.50 R_\mathrm{c}$ in the pure H/He case. In the case shown here, we assume the opacity is constant in the radiative region, $\kappa=0.1$ cm$^2$g$^{-1}$.}
    \label{fig:profile_constant_opacity}
\end{figure*}

In Fig.~\ref{fig:profile_constant_opacity}, we show the structure we find for a planet with a hydrogen-rich atmosphere in equilibrium with the silicate magma ocean below it, such that silicate vapor extends into the atmosphere at the saturation pressure. The magma ocean-atmosphere interface is at 5500~K, and the outer temperature is 1000~K. Coloured lines indicate the radial profiles we calculate for this silicate-containing atmosphere, in (a) pressure $P$, (b) mass mixing ratio of silicate vapor $q_\mathrm{s}$, (c) temperature $T$, (d) lapse rate $\mathrm{d}\log{T}/\mathrm{d}\log{P}$, (e) mean molecular weight $\mu$, and (f) density $\rho$. For comparison, the profiles of an atmosphere composed of H/He with the same mass and planet characteristics but without any silicate vapor are shown as black lines. Solid lines show regions where the atmosphere is convective, while dotted lines show radiative regions. In panel (a), the dashed line shows the partial pressure of silicate vapor, while in (b) the dashed line represents the critical mixing ratio, $q_\mathrm{crit}$, as defined in equation~(\ref{eq:qcrit}), and in (e) the dashed line is the mean molecular weight of the whole atmosphere, $\bar{\mu}$. 

Near the outer radiative-convective boundary (shown by the large dot), the silicate content of the atmosphere is negligible ($q_\mathrm{s} < 10^{-10}$). Therefore, the atmospheric profile is very similar to the convective profile of a pure H/He atmosphere. Some deviation in the lapse rate (panel (d)) occurs with increasing depth due to the latent heat released from condensing silicate vapor (see equation~(\ref{eq:wetadiabat})). However, this deviation is on the order of a $\sim 10$ percent change in lapse rate, so this moist adiabatic profile remains similar to the dry case.

A more substantial change occurs when the mixing ratio reaches the critical mixing ratio, $q_\mathrm{s}=q_\mathrm{crit}$ (panel (b)). At silicate vapor mixing ratios larger than this, convection is inhibited, and the atmosphere becomes radiative. The temperature structure of the inner radiative region is qualitatively different than that of the convective region, but depends sensitively on the opacity in the radiative region, as shown by equation~(\ref{eq:radtemp}). In Fig.~\ref{fig:profile_constant_opacity}, we use a constant opacity, $\kappa=0.1$~cm$^2$g$^{-1}$, and the radiative lapse rate is lower than that of the convective region (panel (d)).

As the temperature increases inward, more and more silicate vapor is stable at equilibrium according to equation~(\ref{eq:Psvp}), resulting in consequent increases in the silicate partial pressure (panel (a)), the mass mixing ratio (panel (b)), and the mean molecular weight (panel (e)). These increases with depth steepen the radial pressure gradient, which scales with mean molecular weight (equation~(\ref{eq:dPdr})). In this case, the steeper radial pressure gradient more than offsets the lower pressure lapse rate, and the overall radial temperature gradient in the radiative region is slightly steeper than in the fully convective case (panel (c)). Therefore, the combination of these temperature and pressure structure effects works to slightly contract the overall radius of the planet compared to a pure H/He case, from about 1.50 to 1.46 core radii. While the atmosphere's mean molecular weight is increased by the silicate vapor to 3.5 amu, the mean molecular weight at the top of the atmosphere remains consistent with pure H/He: none of the silicate vapor appears to be able to reach heights observable via transmission spectroscopy.

\begin{figure*}
\centering
	\includegraphics[width=\textwidth]{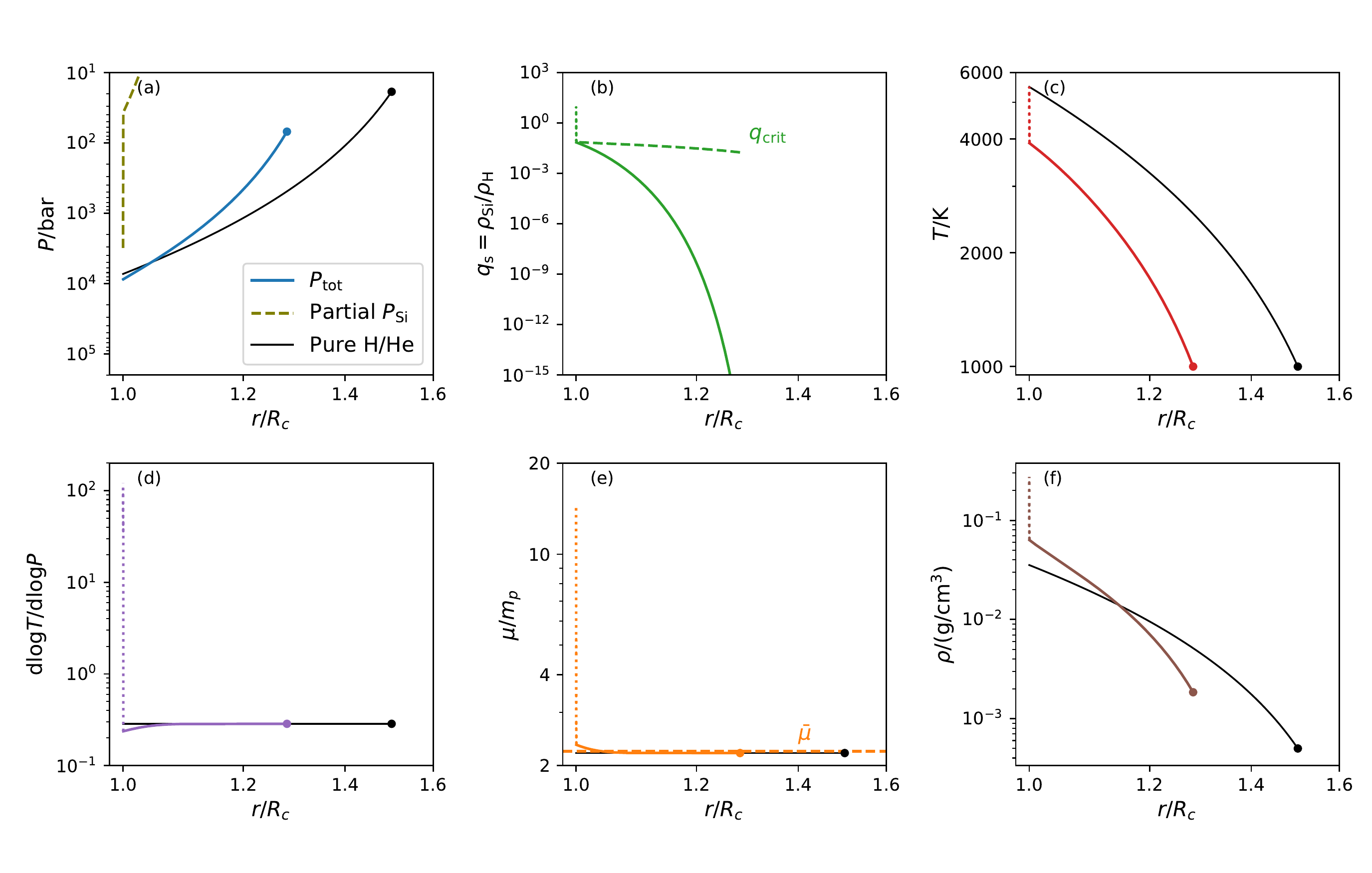}
    \caption{Same as Fig.~\ref{fig:profile_constant_opacity}, but for a radiative region with an opacity that scales with pressure and temperature following the hydrogen opacities of \citet{F14}, who predict much higher opacities than the constant value of 0.1 cm$^2$g$^{-1}$ used in Fig.~\ref{fig:profile_constant_opacity}. Due to this high opacity, the lapse rate becomes very steep in the radiative region, following equation~(\ref{eq:radtemp}), and the region becomes very narrow in width. This significantly decreases the planet's total radius, to $R_\mathrm{rcb} = 1.28 R_\mathrm{c}$, compared to the pure H/He case, which remains at $1.50 R_\mathrm{c}$ as in Fig.~\ref{fig:profile_constant_opacity}.}
    \label{fig:profile_freedman_opacity}
\end{figure*}
As mentioned, while the inhibition of convection is a robust finding, the structure of the radiative region can vary greatly depending on the scaling assumed for the Rosseland mean opacity in the region. In Fig.~\ref{fig:profile_freedman_opacity}, we use the opacity scaling presented for pure H/He in \citet{F14} rather than a constant value. This opacity tends to be much higher than the 0.1 cm$^2$g$^{-1}$ value assumed in Fig.~\ref{fig:profile_constant_opacity} for the high temperature, high pressure conditions at the base of the atmosphere. Since $\partial \ln T/\partial \ln P \propto \kappa$ in the radiative region (equation~(\ref{eq:radtemp})), the radiative gradient becomes very steep, with $\mathrm{d}T/\mathrm{d}P$ more than 100 times larger than that of the moist adiabat (panel (d)). Additionally, the radial pressure gradient increases by a factor of a few due to the increasing mean molecular weight (panel (e)). The combination of these effects, though in this case dominated by the lapse rate effect, leads to a steep radial temperature profile, and the radiative region thus forms nearly a step in temperature at the base of the atmosphere (panel (c)). This narrow radiative region significantly decreases the radius of a planet with the same mass and base temperature: its atmosphere extends essentially the same amount as a fully convective atmosphere with a base temperature equal to the temperature at which the atmosphere transitions from convective to radiative. This acts to decrease the observable radius from the pure H/He atmosphere value of 1.50 core radii to 1.28 core radii, nearly a factor of two in atmospheric width. While the opacity of a gas containing significant silicate vapor may deviate from that of a pure H/He composition, the exact effects are uncertain, so we use the pure H/He scaling from \citet{F14} in the remainder of the work. We discuss the effects of different opacity scalings in Section \ref{sec:opacity}.

To provide further intuitive understanding of the radii we expect for these contracted planets, we analytically approximate the size of an planet with a silicate-induced radiative region in the limit that the radiative region is thin. We can use the fact that at the base of the convective region, $q=q_\mathrm{crit}$. Inserting the definitions of $q$ and $q_\mathrm{crit}$ from section~\ref{sec:silicate_struc}, we obtain
\begin{equation}\label{eq:qqcritP}
    \frac{\mu_\mathrm{v} P_\mathrm{v}}{\mu_\mathrm{H} P_\mathrm{H}} = \cfrac{1}{\bigg(1-\cfrac{\mu_\mathrm{H}}{\mu_\mathrm{sv}}\bigg) \cfrac{\partial \ln P_\mathrm{svp}}{\partial \ln T}} \mathrm{.}
\end{equation}
We convert equation~(\ref{eq:Psvp}) to an exponential functional form of the saturation vapor pressure: $P_\mathrm{v} \equiv \exp{[A-B/T]}$, from which it follows that $\partial \ln P_\mathrm{svp}/\partial \ln T = B/T$. For convenience, we define $\beta \equiv (1-\mu_\mathrm{H}/\mu_\mathrm{sv}) B$.
The hydrogen pressure, $P_\mathrm{H}$, depends on the weight of over-lying atmosphere. Assuming mass is concentrated in the interior of the atmosphere (valid for our choice of $\gamma$), the hydrogen pressure at the base of the atmosphere can be approximated as $P_\mathrm{H} \approx G M_\mathrm{c} M_\mathrm{atm}/(4 \pi R_\mathrm{c}^4)$ \citep[e.g.][]{MS21}. Substituting the temperature dependencies into equation~(\ref{eq:qqcritP}) yields
\begin{equation}\label{eq:qqcritT}
    \frac{\mu_\mathrm{sv}}{\mu_\mathrm{H} P_\mathrm{H}} \exp{\bigg[A-\frac{B}{T}\bigg]} = \frac{T}{\beta} \mathrm{.}
\end{equation}
This equation has no solution for $T$ in terms of elementary functions. However, at the conditions relevant to the sub-Neptunes we study here, the exponential term changes much faster than the linear term. This allows us to approximate the right-hand side as a constant, given some estimate of the temperature we expect. We use an estimate of $T_\mathrm{est} \sim 4500$~K, though the result does not depend strongly on the exact value chosen. Now, we can solve equation~(\ref{eq:qqcritT}) for the temperature at which $q=q_\mathrm{crit}$:
\begin{equation}\label{eq:tempq}
    T_\mathrm{q} = \cfrac{B}{A-\ln\bigg[\cfrac{\mu_\mathrm{H} P_\mathrm{H} T_\mathrm{est}}{\mu_\mathrm{sv} \beta}\bigg]} \mathrm{.}
\end{equation}
We can convert this temperature to a radius using the fact that if the radiative region is concentrated near the surface, the planet is equivalent in size to a fully convective planet with a base temperature given by equation~(\ref{eq:tempq}). To do so, we assume the convective region is dry adiabatic, such that 
\begin{equation}\label{eq:rcbq}
    R_\mathrm{rcb} = \cfrac{R_\mathrm{B}'}{1 + R_\mathrm{B}' /R_\mathrm{c} - T_\mathrm{q}/T_\mathrm{eq}} \mathrm{,}
\end{equation}
where $R_\mathrm{B}' \equiv (\gamma-1)/\gamma \times G M_\mathrm{c} \mu_\mathrm{H}/(k_\mathrm{B} T_\mathrm{eq})$, following \citet{MS21}. This approximation neglects the effects of moist convection on the atmospheric structure. However, since the high temperature portion of the atmosphere is very close to the surface, little silicate vapor extends far from the surface. The mean molecular weight of the entire atmosphere is accordingly nearly the same as for pure H/He (Fig.~\ref{fig:profile_freedman_opacity}, panel (e)). Equation~(\ref{eq:tempq}) yields a temperature at the base of the convective region $T_\mathrm{q}=4013$~K for the parameters of Fig.~\ref{fig:profile_freedman_opacity}. Via Equation~(\ref{eq:rcbq}), this temperature corresponds to a planet radius of $R_\mathrm{rcb}=1.29 R_\mathrm{c}$. Both of these values are in good agreement with the numerical results. This analytic approximation illustrates that the size of these planets is equivalent to that of a planet with a fully convective atmosphere and a base temperature determined by $T_\mathrm{q}$, the temperature at which $q=q_\mathrm{crit}$. This radius can be approximated as solely a function of $M_\mathrm{c}$, $M_\mathrm{atm}$, and $T_\mathrm{eq}$, which may be useful for comparing these findings to observations.

\subsection{Radial evolution}
\begin{figure*}
    \centering
    \includegraphics{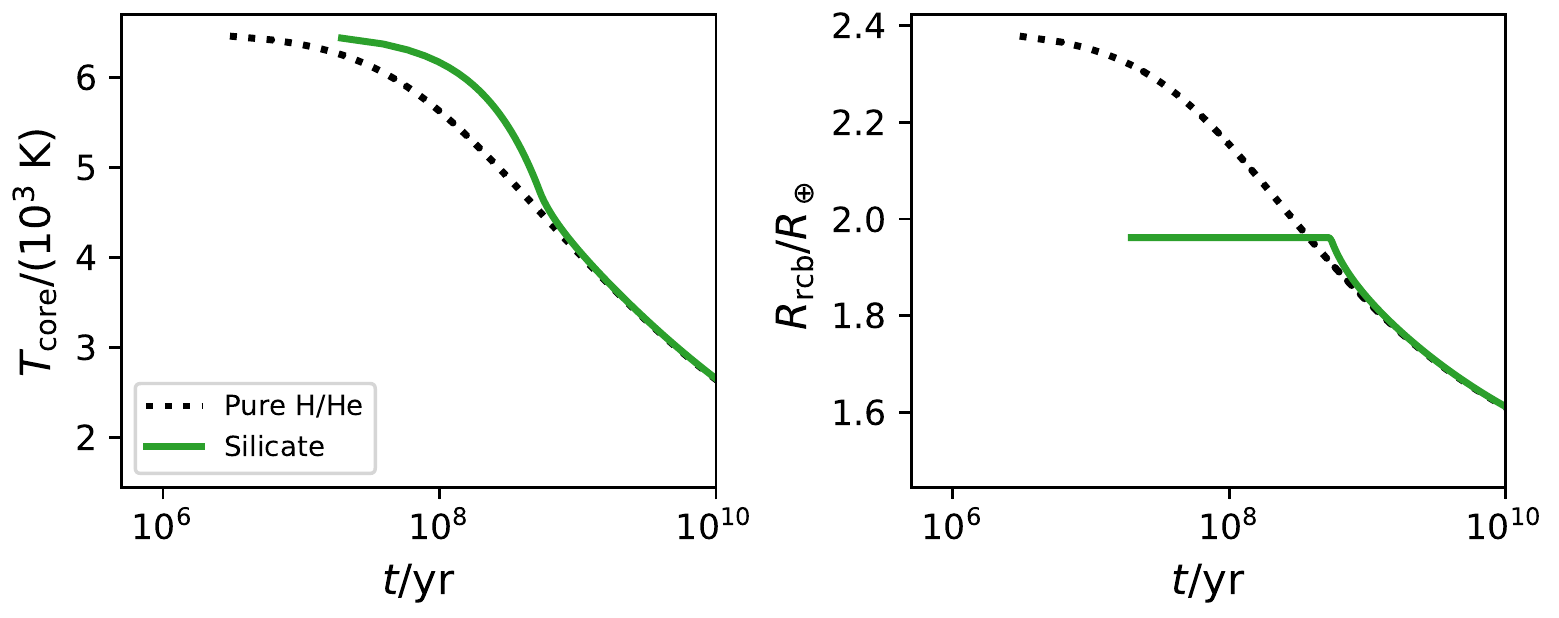}
    \caption{Evolution in temperature at the silicate core-atmosphere interface $T_\mathrm{c}$ (left) and outer radiative-convective boundary $R_\mathrm{rcb}$ (right) of a $4 M_\oplus$ planet with equilibrium temperature $T_\mathrm{eq}=1000$ K. The black dotted line is the time evolution of a pure H/He atmosphere. The green solid line depicts the evolution of an atmosphere containing silicate vapor. For both planets, $t=0$ when the core temperature $T_\mathrm{c}=6500$ K. The silicate vapor atmosphere begins at a much lower radius when the core is hot and a significant radiative layer exists, and its evolution differs on a gigayear timescale.}
    \label{fig:rad_vs_ad}
\end{figure*}

The presence of a radiative region when the planet is hot leads to differences in the evolution of radius and temperature compared to a fully convective atmosphere. We depict these differences in Fig.~\ref{fig:rad_vs_ad}, showing the evolution in core temperature and radiative-convective boundary of two planets. The dotted line represents the evolution of a pure H/He atmosphere with an initial base temperature of 6500~K, while the solid line shows an atmosphere with the same mass and initial conditions containing silicate vapor, with an opacity following \citet{F14}. As described above, when base temperatures are high, the silicate/hydrogen atmosphere is contracted relative to the pure H/He case, with a much lower radiative-convective boundary at the same temperature. This contracted state increases the pressure and density at the radiative-convective boundary, which inhibits cooling, as $L \propto 1/P_\mathrm{rcb}$ by equation~(\ref{eq:luminosity}). Therefore, the silicate-containing atmosphere has a lower initial luminosity than the pure H/He case. Its core temperature consequently goes down more slowly in the early stages of evolution, keeping the atmospheric radius relatively constant as the pure H/He case rapidly contracts.

This slow cooling persists for $\gtrsim 10^8$ years, by which point the radiative-convective boundary of the pure H/He case has decreased to and below that of the silicate/hydrogen case. Now the silicate/hydrogen atmosphere is relatively inflated, with a correspondingly higher luminosity. Therefore, the silicate/hydrogen atmosphere cools relatively rapidly.

A shift in behavior occurs once $T_\mathrm{c} \lesssim 4000$~K. At this point, the mixing ratio at the base of the atmosphere decreases below $q_\mathrm{crit}$. Therefore, the radiative region disappears and the atmosphere becomes fully convective. At this point, the silicate/hydrogen atmosphere can contract as it cools, and rapid contraction decreases its radius until it is of similar size to the pure H/He case. From this point on, the radial and thermal evolution of the two atmospheres are very similar.

\subsection{Population statistics}
\begin{figure*}
    \centering
	\includegraphics[width=0.45\textwidth]{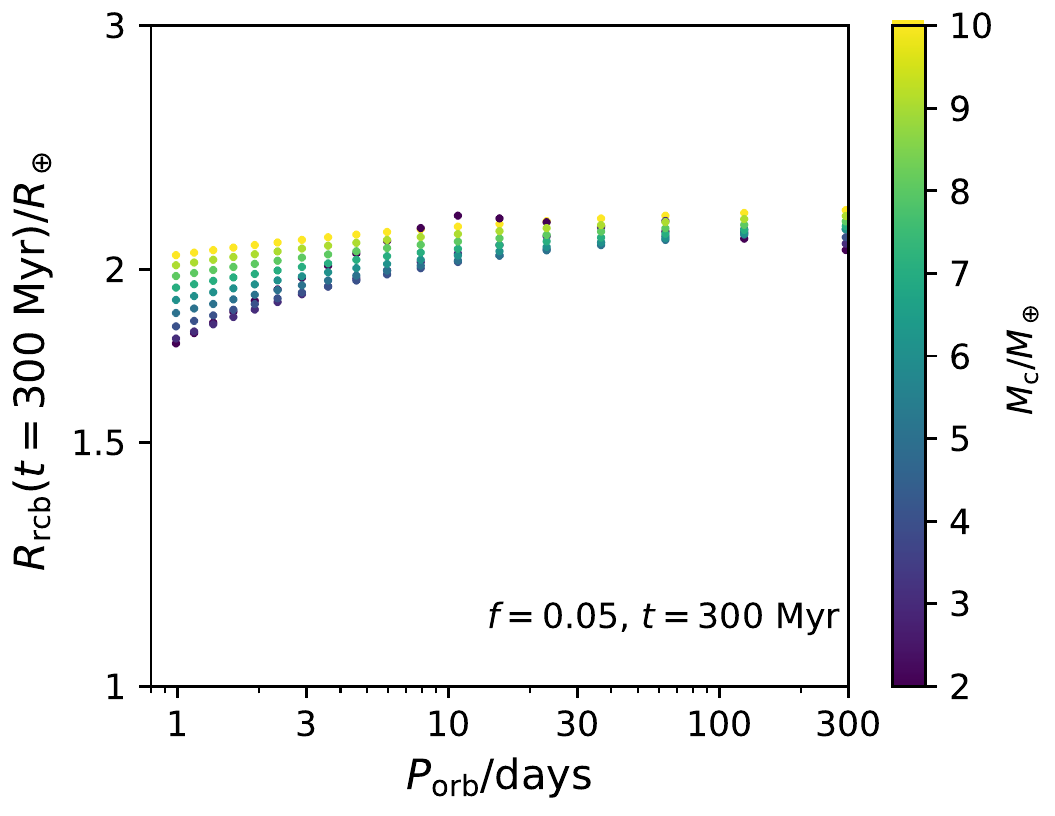}
	\includegraphics[width=0.45\textwidth]{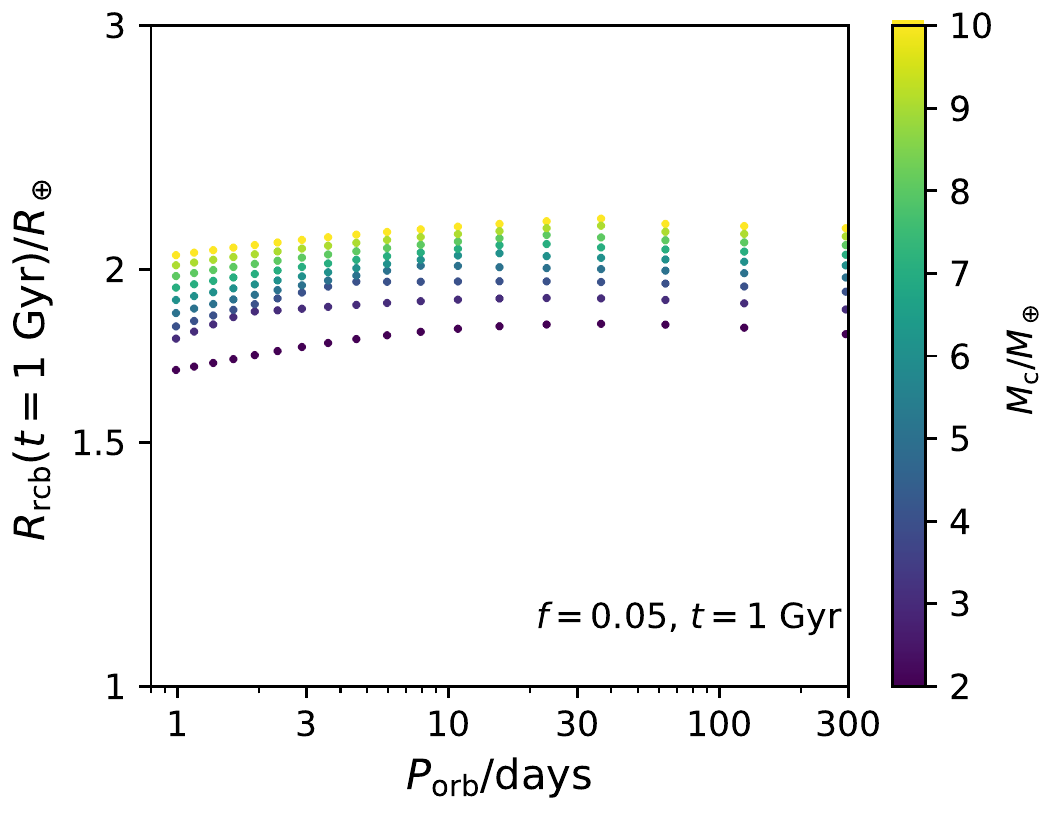}
    \caption{Planet radius evolution when silicate vapor is included in the atmosphere. Plotted are the outer radiative-convective boundary radius as a function of orbital period around a Sun-like star, for different planet masses denoted by colors. The left panel shows the radii at 300~Myr after the planets had base temperatures of 6500~K, while the right panel shows the radii after 1~Gyr. Cooler planets at longer orbital periods have larger radii for the same initial base temperature but cool more quickly, making intermediate temperature planets the largest for a given mass.}
    \label{fig:rcb_evol}
\end{figure*}

These differences in atmospheric structure and evolution between planets with pure H/He and silicate/hydrogen atmospheres manifest on a population level, affecting the radii planets have and how they depend on attributes like core mass, atmospheric mass, and equilibrium temperature. In Fig.~\ref{fig:rcb_evol}, we show the radii of a suite of planets with varying orbital periods. The color gradient of the dots represents different core masses. On the left panel, we show the radii at $t=300$~Myr since the planets had core temperatures of 6500~K, whereas on the right panel we show the same planets after 1~Gyr of thermal evolution. All these planets have the same atmospheric mass fraction, $f=0.05$.

The presence of a silicate/hydrogen atmosphere does not affect these planets equally. Fig.~\ref{fig:rcb_evol} demonstrates that while more massive planets still tend to have larger radii than less massive planets, the difference is less, and in some cases the less massive planets have larger radii than those more massive. This is due to their weaker gravity leading to more inflated atmospheres for planets with the same base temperature and atmospheric mass fraction. However, these more extensive envelopes lead to larger luminosities and more rapid cooling. Therefore, planets at longer orbital periods become fully convective and contract more rapidly than shorter-period planets. This leads to a non-monotonic scaling of radiative-convective boundary at a given time with orbital period: intermediate period planets have larger radii than those with shorter or longer periods. The planet with the largest radius is the coolest planet that has not yet begun to contract. At 300~Myr, this occurs at periods of about 10~days for $M_\mathrm{c}=2 M_\oplus$, whereas the more massive planets all remain in the radiative region for $P>300$~days. At 1~Gyr, the $2 M_\oplus$ planets have all contracted, while the peak radius is at hotter temperatures for less massive planets.

\subsubsection{Comparison with H/He only models}
\begin{figure*}
    \centering
	\includegraphics[width=0.32\textwidth]{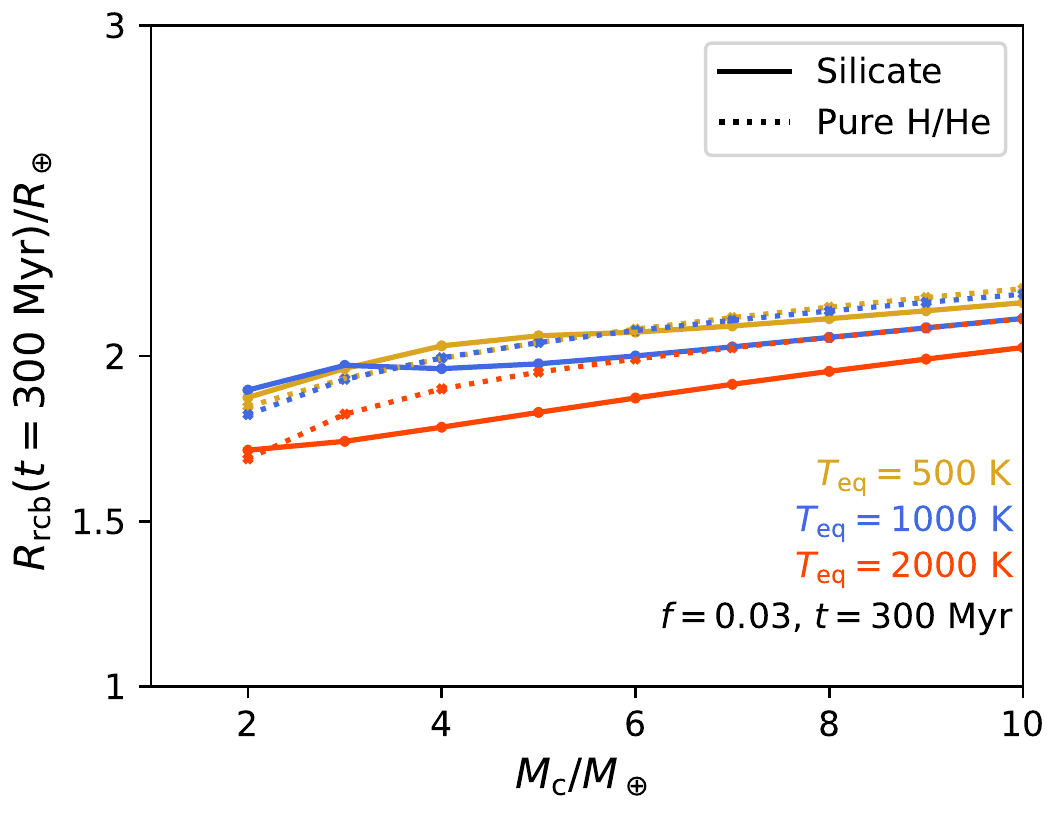}
	\includegraphics[width=0.32\textwidth]{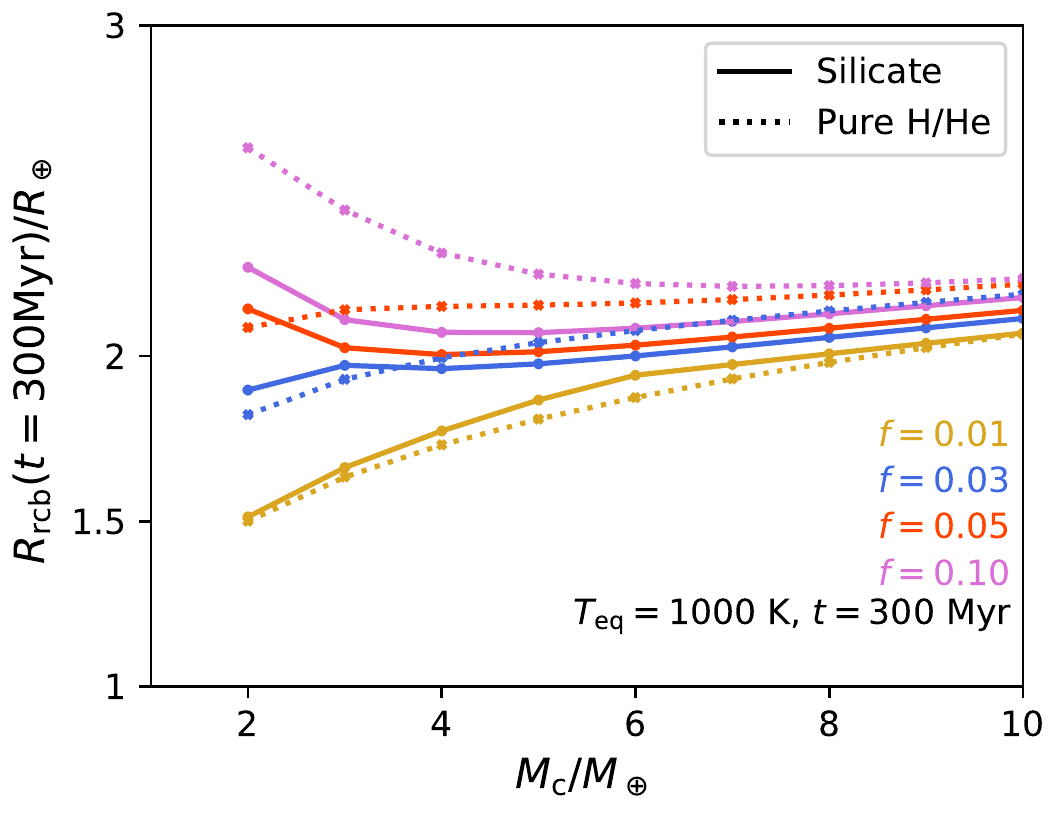}
	\includegraphics[width=0.32\textwidth]{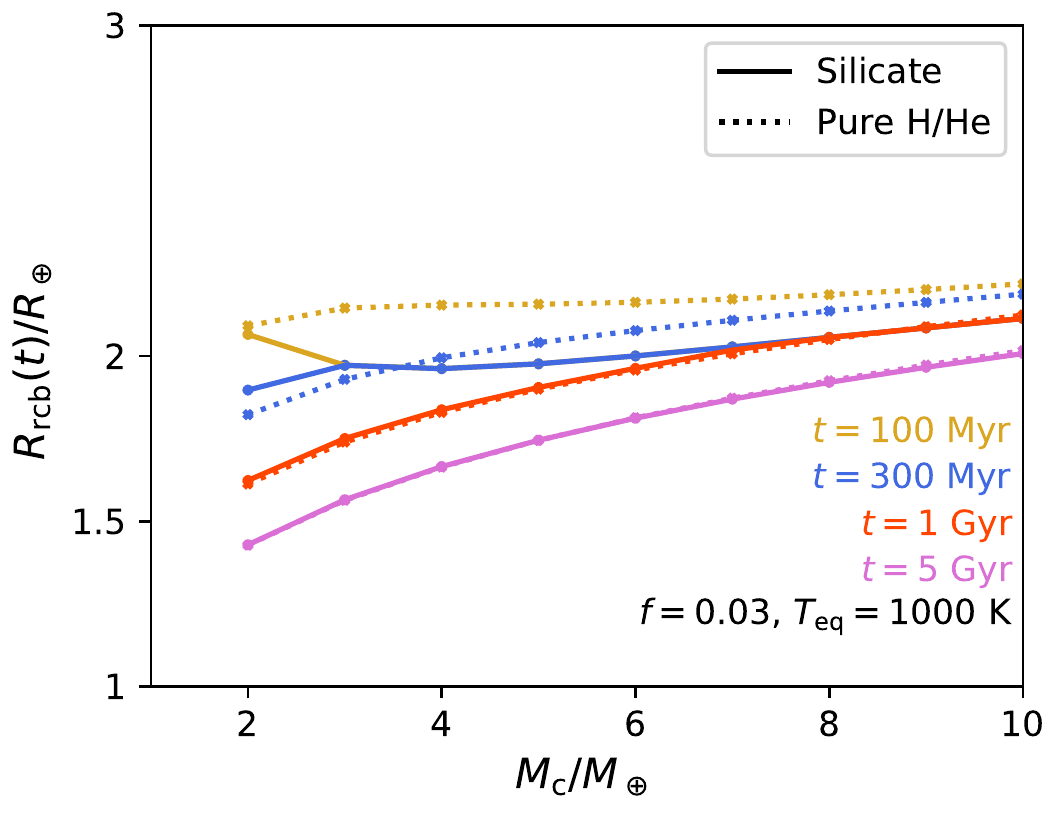}
    \caption{Comparison of the evolution of planet radius if silicate vapor condensation in the atmosphere is considered (solid lines) to a model with a convecting pure H/He atmosphere (dotted lines) as a function of planet mass. Colors represent (left) different equilibrium temperatures, (center) different atmospheric mass fractions, and (right) different times since the initial base temperature.}
    \label{fig:M_vs_Rrcb}
\end{figure*}

One major implication of these results is that the radial evolution of sub-Neptune planets may be substantially different if the silicate-induced radiative layer is accounted for than in a model assuming a pure H/He atmosphere, which is fully convective between $R_\mathrm{c}$ and $R_\mathrm{rcb}$. In Fig.~\ref{fig:M_vs_Rrcb} we plot the radiative-convective boundary radius for a pure H/He model, shown by dotted lines, and our hydrogen/silicate model, shown by solid lines, as a function of core mass. In all cases, evolution begins, i.e. $t=0$, when the temperature at the base is $6500~$K, and the times provided are times since this the planet had this initial base temperature. The colors represent different equilibrium temperatures (left), atmospheric mass fractions (center), and times since $T_\mathrm{c}=6500~$K.

In the left panel, we compare the radii of three sets of planets at a constant time, $t=300$~Myr, and atmospheric mass fraction, $f=0.03$. The different colors denote equilibrium temperatures of 500, 1000, and 2000~K. Hotter, more massive planets in models which include silicate vapor tend to have smaller radiative-convective boundaries relative to the pure H/He case, as these atmospheres are still hot enough at their bases to maintain a radiative region. Lower mass and cooler planets with hydrogen/silicate atmospheres, on the other hand, can be comparable in size or even inflated relative to the pure H/He case, as their atmospheres have cooled sufficiently such that the radiative region is no longer present. Therefore, these planets have begun to contract.

In the center panel, we similarly compare four sets of planets, this time varying the atmospheric mass fraction between 1 and 10 percent of a planet's mass. The silicate/hydrogen case is the most contracted relative to the H/He case when the atmospheric mass fraction is highest, since more massive atmospheres cool more slowly. Meanwhile, rapid evolution leads to less massive Si/H atmospheres being comparable in size to H atmospheres, or even slightly inflated.

Finally, on the right panel we compare planets in time. For $f=0.03$ and $T_\mathrm{eq}=1000$~K, the more massive planets maintain silicate-induced radiative layers, preventing significant radius change, for $>1$~Gyr. However, the corresponding H/He-only atmosphere cools and contracts in this time, approaching the silicate/hydrogen atmosphere's radius. Therefore, the silicate/hydrogen atmosphere evolves from relatively contracted at early times to similar in extent to the hydrogen-only case at 1~Gyr. Less massive planets evolve on shorter timescales: the $4 M_\oplus$ planet with a Si/H atmosphere is only relatively contracted for $<300$~Myr, while a $2 M_\oplus$ planet with a Si/H atmosphere is already contracting at $100$~Myr.

\begin{figure*}
    \centering
	\includegraphics[width=0.32\textwidth]{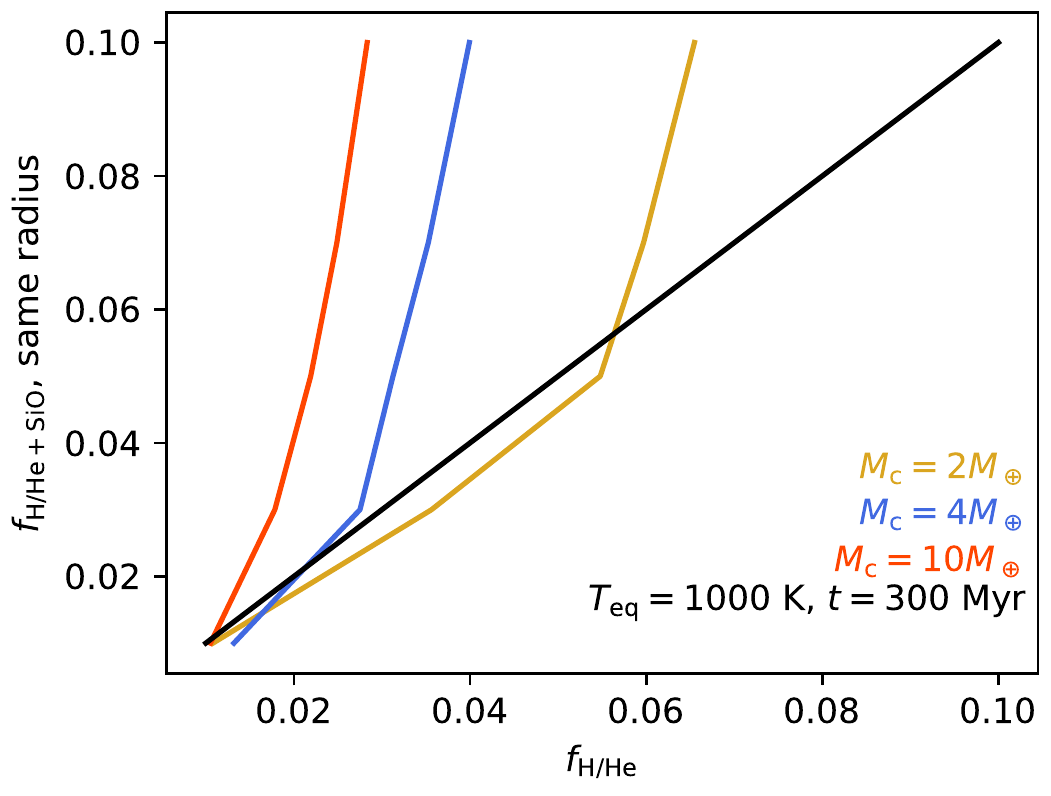}
	\includegraphics[width=0.32\textwidth]{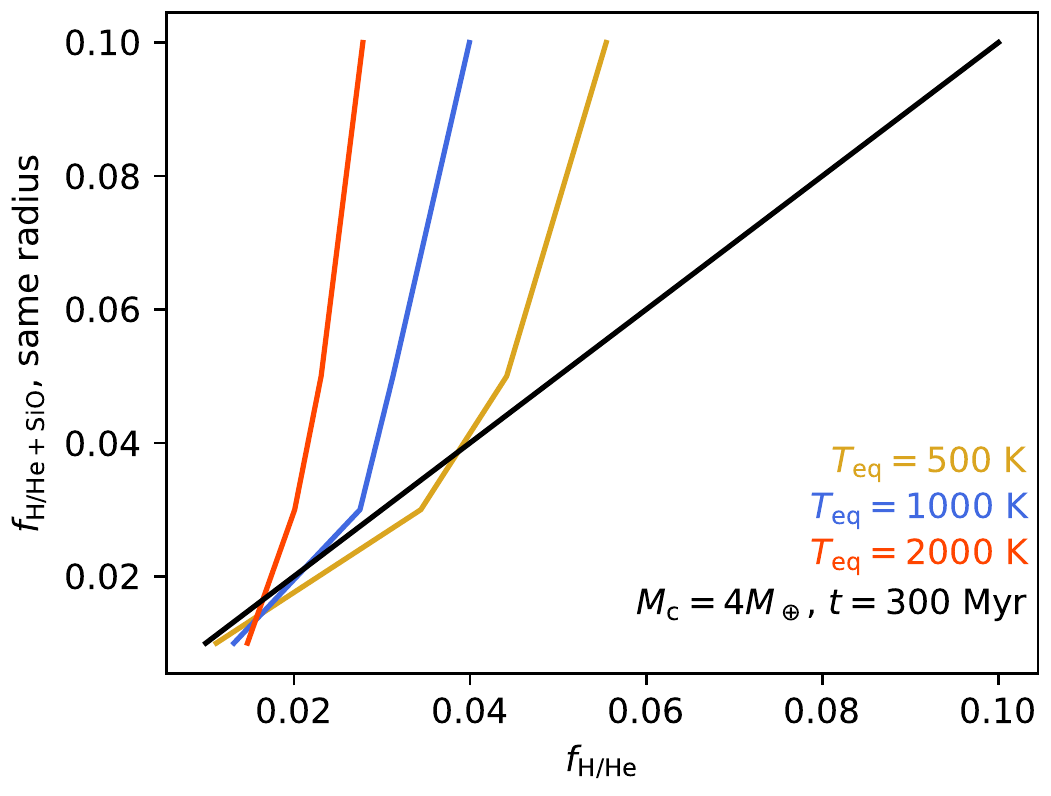}
	\includegraphics[width=0.32\textwidth]{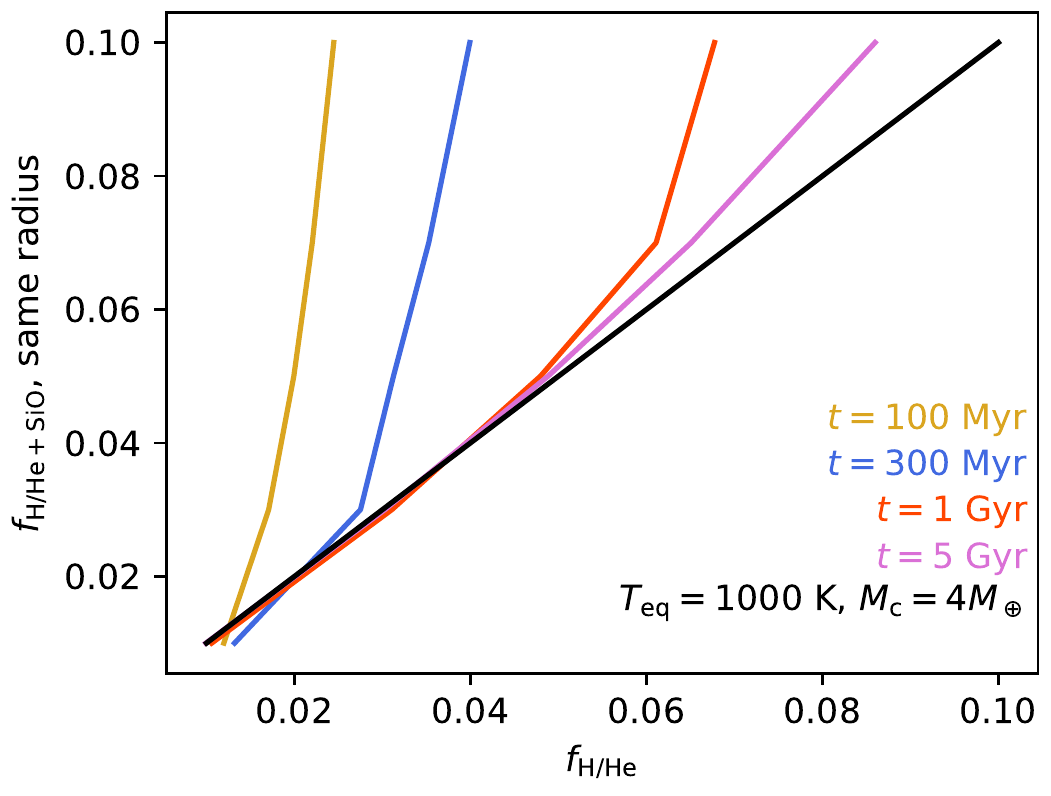}
    \caption{Comparison of atmospheric mass fraction inferred assuming the atmosphere is composed of pure H/He which is fully convecting to a model which includes a silicate-induced radiative layer with the same outer radiative-convective boundary radius. On the left is the trend in inferred mass fraction for three different planet masses at the same time and equilibrium temperature. The center panel compares planets of fixed mass at three equilibrium temperatures at a given time. The right-most panel shows the same planet mass and equilibrium temperature at four different times. Young planets close to their stars with large core and atmospheric masses tend to cool relatively slowly and thus retain a radiative layer at the base of the atmosphere for longer. This radiative layer decreases their radius compared to the pure H/He case, meaning the same radius corresponds to a larger atmospheric mass. Conversely, low mass, low $f$, and older planets with low equilibrium temperatures cool quickly and lose their inner radiative layers. These planets therefore have similar radii in both models, and so come close to lying along the one-to-one line in black. Atmospheric models which do not include silicate condensation could underestimate the masses of sub-Neptune planets by a factor of 5 for the most extreme cases shown here.}
    \label{fig:f_inferred}
\end{figure*}

Evolution models conceptually similar to those presented above are typically used to infer the atmospheric mass for exoplanets with constrained planet radii, masses, ages, and equilibrium temperatures. However, such models typically assume a fully convective hydrogen/helium atmosphere, neglecting the silicate vapor we have demonstrated to be important to the structure and evolution of these planets \citep[e.g.][]{LF14}. In Fig.~\ref{fig:f_inferred}, we compare the atmospheric mass inferred using a purely convective structure to that which we obtain from our models which include SiO in the atmosphere for the same planet radius. 

On the left we show a comparison for three different planet masses with fixed $t$ and $T_\mathrm{eq}$. Generally, most planets for which silicate vapor is included have higher inferred atmospheric masses than those inferred for the same radii assuming a pure H/He, convective structure. The black line shows a one-to-one relationship for reference. This is most apparent at high atmospheric masses and high planet masses: a $2 M_\oplus$ planet with $f=0.10$ modeled to include silicate vapor has a radius equivalent to that of a pure H/He planet with $f \approx 0.06$, and a $10 M_\oplus$ planet with $f=0.10$ has a radius for which one would infer $f \approx 0.02$ if a pure H/He composition was assumed. For lower atmospheric mass fractions, the inferred atmospheric masses are comparable between the two models. This is because lower mass atmospheres have lower densities, and therefore cool more quickly per equation~(\ref{eq:luminosity}). This relatively fast cooling leads to the silicate/hydrogen atmospheres becoming fully convective sooner, and their radii therefore approach the radii of planets with equal mass pure H/He atmospheres. Some silicate-containing atmospheres, such as the $2 M_\oplus$ planets with intermediate atmospheric masses, $f=0.03$ or 0.05, are actually slightly larger than pure hydrogen atmospheres at the same time since $T_\mathrm{c}=6500$~K and therefore have lower inferred atmospheric masses. At this time, the planets with silicate-induced radiative regions remain at their initial fixed radius, but the hydrogen-only atmospheres have contracted to radii smaller than this. Such an effect is visible in Fig.~\ref{fig:rad_vs_ad} at a few hundred Myr. This relatively inflated state does not last long, as these planets now cool quickly relative to the more contracted pure H/He atmospheres. This cooling allows the formerly-inflated planets to become fully convective and shrink, catching up to their H/He-only counterparts. The overall effect of this inflated state is minor compared to the difference in radii at earlier times.

Similarly to the left panel, the center panel compares three equilibrium temperatures at fixed $t$ and $M_\mathrm{c}$, while on the right we show planets with the same core mass and equilibrium temperature at four different ages. Increasing equilibrium temperature inhibits cooling. Hence, a hotter planet will exhibit a larger discrepancy in inferred envelope mass between the silicate/hydrogen and pure H/He models than a cooler planet. 
This is illustrated in Fig.~\ref{fig:f_inferred}: for fixed core mass, $M_\mathrm{c}=4 M_\oplus$, and time, $t=300$~Myr, a close-in planet with $T_\mathrm{eq}=2000$~K and a silicate/hydrogen atmosphere of mass $f=0.10$ is the same size as a planet with a H/He-only atmosphere with one fifth the atmospheric mass ($f \approx 0.02$). But an equal-mass atmosphere around a planet with $T_\mathrm{eq}=500$~K matches the size of a pure-H/He atmosphere only a factor of 2 smaller in mass ($f \approx 0.05$). Meanwhile, as planets age, they cool sufficiently such that the atmosphere becomes fully convective. This allows the silicate/hydrogen atmospheres to contract as they cool, narrowing the difference between the extents of silicate/hydrogen and pure hydrogen atmospheres. For example, for $T_\mathrm{eq}=1000$~K and $M_\mathrm{c}=4 M_\oplus$, at 100~Myr after $T_\mathrm{c}=6500$~K, there are significant differences between these two models for $f>0.01$. At this time, consideration of silicate condensation produces at least a factor of 2 difference in inferred mass for $f=0.03$, up to a factor of 5 for larger atmospheric mass fractions. By $t=1$~Gyr, the lower atmospheric mass planets with silicate/hydrogen atmospheres have become indistinguishable from the pure H/He atmospheres, but differences persist for $f>0.05$. These differences become smaller with time, as shown by the contours approaching the one-to-one black line, but discrepancies of 20 percent between the modeled masses persist even at 5~Gyr for the most massive atmospheres we consider.

To summarize, we find that the silicate-induced radiative region causes H/He-only models to under-estimate the true atmospheric mass of sub-Neptunes. The mass inferred is most disparate for planets with higher core mass, higher atmospheric mass, higher equilibrium temperatures, and younger ages. This follows from the relatively contracted radii of planets with hot interiors when silicate vapor is considered. As these planets cool, their radiative regions disappear, and they begin contracting. However, in certain cases this contraction does not occur until the planet is larger than a pure H/He planet at the same age. In this case, the silicate/hydrogen atmosphere will appear inflated relative to the H/He case, and the mass inferred will consequently be smaller, causing excursions below the one-to-one line in Fig.~\ref{fig:f_inferred}. Over time, planets approach the radius one would expect for a pure H/He atmosphere. Differences can persist for planets with the largest initial differences for $>5$~Gyr. These differences in radii, which translate to a factor of $\sim 5$ difference in inferred atmospheric mass for the youngest, most massive, or hottest planets we consider here, indicate there could be substantial error in inferred hydrogen mass fractions for observed exoplanets when the effects of silicate vapor are not considered.

\section{Discussion and Future Work}\label{sec:discussion}
In the section above, we have shown that silicate-induced radiative regions deep in the interiors of sub-Neptune atmospheres can have a large effect on the size, and therefore the inferred atmospheric mass fraction, of this common class of exoplanets. Below we discuss the effects of some uncertainties in our model. These uncertainties include the opacity behavior in the deep atmosphere, the initial base temperature with which planets begin their evolution, and the effect of condensables besides silicate vapor. We also outline prospects for future work, including the integration of these novel atmospheric structures with models of atmospheric mass loss.

\subsection{Opacities}\label{sec:opacity}
In Sec.~\ref{sec:results}, we demonstrated that the opacity structure has a large effect on the extent of the radiative region. A low, constant opacity such as 0.1~cm$^2$g$^{-1}$ leads to a slow increase in temperature with increasing pressure. Consequently, the radiative layer becomes thick, extending out to one third of the atmosphere's width in Fig.~\ref{fig:profile_constant_opacity}. Conversely, an opacity that scales with density as in \citet{F14} leads to larger opacity values in the interior, typically exceeding 1000~cm$^2$g$^{-1}$. These large opacities produce a very steep temperature gradient in the interior, narrowing the radiative region to a small fraction of the planet's total radius, as depicted in Fig.~\ref{fig:profile_freedman_opacity}. While there are many theorized opacity scalings, which likely depend on the exact silicate content of the atmosphere \citep[e.g][]{F14, LCO14}, the steepness of the temperature gradient in the high opacity case makes our results insensitive to the exact opacity scaling used. So long as the opacity is sufficiently high such that the width of the radiative region becomes much less than the total width of the atmosphere, the exact structure has little effect on observables such as the planet's radius and inferred atmospheric mass. This is because the planet's radius is dominated by the structure of the convective region, and therefore the temperature at which $q=q_\mathrm{crit}$, which is independent of the opacity of the radiative layer.

We can quantify the effect of the opacity on the thickness of the atmosphere by combining hydrostatic equilibrium (equation~(\ref{eq:dPdr})) and the radiative lapse rate (equation~(\ref{eq:radtemp})) to solve for the width of the radiative region, $\Delta R$, given the change in temperature across it, $\Delta T$. We find that
\begin{equation}\label{eq:opacity_radius}
\begin{split}
    \frac{\Delta R}{R_0} &= \frac{64 \pi}{3} R_0 \frac{k_\mathrm{B} \Delta T}{\mu} \frac{\sigma T_0^4}{\kappa L P_0} \\
    &\approx 0.01 \bigg(\frac{R_0}{10^9~\mathrm{cm}}\bigg) \bigg(\frac{\Delta T}{2000~\mathrm{K}}\bigg) \bigg(\frac{T_0}{4000~\mathrm{K}}\bigg)^{4} \bigg(\frac{\kappa}{10~\mathrm{cm^2 g^{-1}}}\bigg)^{-1} \\ &\bigg(\frac{L}{10^{22}~\mathrm{erg s^{-1}}}\bigg)^{-1} \bigg(\frac{P_0}{10^5~\mathrm{bar}}\bigg)^{-1}
\end{split}
\end{equation}
where $R_0$, $T_0$, and $P_0$ represent the radius, temperature, and pressure at the radiative-convective transition. Specifically, in the thin radiative region limit, $R_0 \approx R_\mathrm{c}$. In the second equality, we scale this equation using typical values for the super-Earths we consider. Using these typical values, equation~(\ref{eq:opacity_radius}) shows that if the opacity $\kappa>10$~cm$^2$g$^{-1}$, the width of the radiative region $\Delta R < 0.01 R_0$, negligibly thin compared to the width of the whole atmosphere. This opacity condition is easily met by extrapolations of the opacity laws described in, e.g., \citet{F14} and \citet{LCO14}. In summary, while the exact opacity prescription best-suited to a hydrogen atmosphere containing substantial silicate vapor may be uncertain, the effects of alternative opacity scalings on atmospheric observables will be negligible so long as the opacity remains sufficiently high.

\subsection{Conduction}
At the typical temperatures and pressures in the deep non-convective regions of sub-Neptunes we consider in this work, conduction may be competitive with radiation in transporting heat \citep[e.g.][]{Vazan2020}. As with the opacities discussed in the preceding paragraph, the thermal conductivities in these regions, $\lambda$, are uncertain and depend on the material properties of the atmosphere. Under sufficiently high pressure and temperature, hydrogen gas becomes metallic, which frees electrons that can easily transport thermal energy and thus leads to high conductivities. However, \textit{ab initio} calculations and experimental evidence indicate this transition mostly takes place at higher temperatures and pressures than we consider \citep[e.g.][]{FrenchBecker2012,McWilliams2016}. In the sub-Neptune regime, thermal conductivities are typically on the order of $10^5$ to $10^6$ erg s$^{-1}$ cm$^{-1}$ K$^{-1}$, i.e., 1 to 10 W m$^{-1}$ K$^{-1}$ \citep[e.g.][]{McWilliams2016}. These values for hydrogen are similar to the thermal conductivity typically used for Earth-like silicate in planetary interiors, $\lambda \sim 4$ W m$^{-1}$ K$^{-1}$ \citep[e.g][]{Stevenson1983,Vazan2020}, and that derived from \textit{ab initio} simulations of silicate liquids at planetary conditions \citep{SSD17}. Using these conductivity values, we find that conductive heat transport can be similar in magnitude to radiation for the conditions considered here but does not significantly exceed it.

To verify these results, we modeled the atmospheric structure including both conduction and radiation. For the thermal conductivities, we use the electrical conductivity scaling of \citet{McWilliams2016}, based on experimental results for pure hydrogen, and convert to thermal conductivities using the Wiedemann-Franz law. At lower temperatures and pressures, where the electrical conductivity is low, we use a minimum value of $\lambda = 2 \times 10^5$ erg s$^{-1}$ cm$^{-1}$ K$^{-1}$, appropriate for the nucleic contribution over a broad range of relevant temperatures and pressures \citep{FrenchBecker2012}. We found that the overall atmospheric radii were not measurably different when both conduction and radiation were included than the previous, radiation-only results presented in Section~\ref{sec:results}. In both cases, the widths of the non-convective regions were negligibly thin compared to the overall planet radius, even for the most massive planets and atmospheres we consider in this work. These results support the conclusion that while conduction could be competitive with radiation deep in some of these atmospheres, it does not affect the overall qualitative and quantitative findings of this work for the temperatures and pressures we consider.

\subsection{Initial base temperature}
The initial base temperatures chosen throughout this work are in the range suggested by accretion models \citep[e.g.][]{GSS16}, but it is possible the initial temperature at the base of sub-Neptunes could be higher, up to 10000~K. An increased starting base temperature results in more core thermal energy available for cooling. However, the initial radius is virtually unchanged, due to the steepness of the radiative region. Therefore, the cooling timescale becomes longer, prolonging evolution and contraction to longer timescales than presented in Sec.~\ref{sec:results}. The qualitative effect of silicate vapor decreasing the radius, and therefore increasing the inferred atmospheric mass, of sub-Neptune planets holds so long as the initial base temperature is $\gtrsim 4000$~K, i.e., high enough for a radiative region to form.

Additionally, this initial base temperature may vary from planet to planet a function of a planet's physical parameters, as more massive planets may begin with hotter base temperatures. This effect would magnify the increase in cooling timescales already present with increasing planet mass. Determining these initial conditions in detail requires modeling the formation of these planets and their H/He accretion from the protoplanetary disk, which is beyond the scope of this work. These higher temperatures also go beyond the temperature range for which the silicate vapor pressure relation we employ \citep{VF13} was originally intended.

\subsection{Other condensables}\label{sec:cond}
While equilibrium chemistry models find silicate vapor to be by far the most abundant vapor species in equilibrium at the magma ocean-atmosphere interface of sub-Neptunes \citep{SY21}, other species are expected to be present in lower concentrations. One such species is water vapor, which may be present at the $\sim 10$ percent level above an Earth-composition magma ocean. Water vapor abundance constrains both habitability and planet migration in early stellar systems, so accurately quantifying the endogenic water concentrations we expect from sub-Neptunes is important. Enhanced atmospheric abundances of water are possible if sub-Neptunes have a more ice-rich composition than Earth, which are also consistent with measured bulk densities \citep[e.g.][]{RS10,D17,Zeng2019,Mousis2020}. If sub-Neptunes are water-rich, high-pressure ice layers could form between the silicate core and H/He atmosphere \citep[e.g.][]{NixonMadhusudhan2021}, which could impede the interaction between the silicate and H/He investigated in this work. Additionally, the endogenic production of water vapor could affect the quantity of SiO vapor expected above a magma ocean \citep{SY21}, behavior which we aim to more fully capture in future work.

Since water vapor is expected to be at lower concentrations than silicate vapor in equilibrium with a magma ocean \citep{SY21}, we expect its importance in influencing atmospheric structure to be of lower order. However, since water condenses at a much lower temperature than silicate vapor, whatever water is at equilibrium at the base of the atmosphere may extend into regions observable by transmission spectroscopy. Additionally, this lower condensation temperature could mean that this endogenic water's effects could impact a different region of the atmosphere than the silicate vapor, even if the magnitude is smaller. However, this region of maximum effect may be at a temperature lower than $T_\mathrm{eq}$ for some of the sub-Neptunes we consider here, so water's effect on the atmospheric structure may be small for these hotter planets.

\subsection{Mass loss processes}
Besides the observable consequences of different planet radii, the inhibition of convection by silicate vapor atmospheres could also affect the mass loss processes thought to shape the radius valley, such as photo-evaporation \citep[e.g.][]{OW17} and core-powered mass loss \citep[e.g.][]{GS19}. The implications of silicate-induced radiative regions on the accretion of hydrogen gas from the protoplanetary disk and subsequent hydrodynamic mass loss should be carefully considered. We intend to integrate these novel silicate-induced structures into models of mass loss processes in future work.

\section{Summary and Conclusions}\label{sec:conc}
In this work, we have demonstrated that silicate vapor has significant effects on the structure of sub-Neptune atmospheres. At high temperatures, condensation of silicate vapor in a hydrogen-rich atmosphere can induce a mean molecular weight gradient that inhibits convection, leading to a radiative atmospheric profile near the magma ocean-atmosphere interface of these planets. The exact temperature gradient depends on the opacity dependence of the atmosphere, but for opacities typical of high-density hydrogen, the gradient is very steep. Therefore, the temperature drops sharply above the magma ocean until the silicate abundance is low enough to allow for convection. This radiative layer decreases the overall radius of a planet compared to a fully convective, pure H/He atmosphere with the same base temperature. We simulate the thermal evolution in time of these planets and find that young planets with silicate/hydrogen atmospheres and base temperatures $\gtrsim 4000$~K are much smaller than pure H/He atmospheres with the same base temperatures. As these planets cool, all the change in temperature at the base is accommodated by the radiative region, preventing contraction: a planet with any inner radiative region has nearly constant radius no matter the base temperature. This shrunken state leads to slower cooling, allowing equivalent convective H/He atmospheres to cool and contract until they approach and are briefly smaller than the silicate/hydrogen atmospheres. Once the abundance of silicate vapor is low enough, the planet can contract, and eventually the two cases converge. This convergence happens more slowly for planets with larger masses and atmospheric mass fractions, and silicate/hydrogen atmospheres can have substantially different inferred atmospheric masses on gigayear timescales.

We survey the sub-Neptune parameter space to quantify how these differences in evolution depend on a planet's physical parameters. We find that high mass planets with high atmospheric mass fractions differ most substantially at any given time. While lower mass planets with smaller atmospheric mass fractions also have large radius differences, these planets also cool more quickly, thereby losing their radiative regions and erasing the initial dichotomy. Meanwhile, while planets at cooler equilibrium temperatures start the most inflated, they also cool the fastest, leading to intermediate temperature planets being the largest at any given time. Finally, we compare the atmospheric masses inferred from a pure H/He atmospheric model to those inferred from the same radius in a silicate/hydrogen atmospheric model. We find that for a $10 M_{\oplus}$ planet with an equilibrium temperature of 1000~K, a hydrogen-dominated atmosphere containing silicate vapor with $f=0.10$ has the same radius as hydrogen-only model with $f \approx 0.02$, if both planets have cooled for 300~Myr from an initial temperature at the bases of their atmospheres of 6500~K. In essence, the silicate-induced radiative layer typically makes the atmosphere more contracted than it would be if the atmosphere were pure H/He. Such differences can persist for gigayears, especially for more massive planets with larger atmospheric mass fractions. Therefore, atmospheric masses inferred from measured exoplanet radii can be substantially under-estimated if compositional equilibrium with the underlying silicates is not considered.

\section*{Acknowledgements}

We thank the anonymous reviewer and Allona Vazan for insightful comments which improved the manuscript. In this work we use the \textsc{numpy} \citep{numpy}, \textsc{matplotlib} \citep{Matplotlib}, and \textsc{scipy} \citep{scipy} packages. This research has been supported in part by the National Aeronautics and Space Administration under grant No. 80NSSC18K0368 issued through the Exoplanet Research Program.
%%%%%%%%%%%%%%%%%%%%%%%%%%%%%%%%%%%%%%%%%%%%%%%%%%
\section*{Data Availability}
Data available on request.

%%%%%%%%%%%%%%%%%%%% REFERENCES %%%%%%%%%%%%%%%%%%

% The best way to enter references is to use BibTeX:

\bibliographystyle{mnras}
\bibliography{si_vapor} % if your bibtex file is called example.bib

\begin{thebibliography}{}
\makeatletter
\relax
\def\mn@urlcharsother{\let\do\@makeother \do\$\do\&\do\#\do\^\do\_\do\%\do\~}
\def\mn@doi{\begingroup\mn@urlcharsother \@ifnextchar [ {\mn@doi@}
  {\mn@doi@[]}}
\def\mn@doi@[#1]#2{\def\@tempa{#1}\ifx\@tempa\@empty \href
  {http://dx.doi.org/#2} {doi:#2}\else \href {http://dx.doi.org/#2} {#1}\fi
  \endgroup}
\def\mn@eprint#1#2{\mn@eprint@#1:#2::\@nil}
\def\mn@eprint@arXiv#1{\href {http://arxiv.org/abs/#1} {{\tt arXiv:#1}}}
\def\mn@eprint@dblp#1{\href {http://dblp.uni-trier.de/rec/bibtex/#1.xml}
  {dblp:#1}}
\def\mn@eprint@#1:#2:#3:#4\@nil{\def\@tempa {#1}\def\@tempb {#2}\def\@tempc
  {#3}\ifx \@tempc \@empty \let \@tempc \@tempb \let \@tempb \@tempa \fi \ifx
  \@tempb \@empty \def\@tempb {arXiv}\fi \@ifundefined
  {mn@eprint@\@tempb}{\@tempb:\@tempc}{\expandafter \expandafter \csname
  mn@eprint@\@tempb\endcsname \expandafter{\@tempc}}}

\bibitem[\protect\citeauthoryear{{Biersteker} \& {Schlichting}}{{Biersteker} \&
  {Schlichting}}{2021}]{BS21}
{Biersteker} J.~B.,  {Schlichting} H.~E.,  2021, \mn@doi [\mnras]
  {10.1093/mnras/staa3614}, \href
  {https://ui.adsabs.harvard.edu/abs/2021MNRAS.501..587B} {501, 587}

\bibitem[\protect\citeauthoryear{{Brouwers} \& {Ormel}}{{Brouwers} \&
  {Ormel}}{2020}]{BrouwersOrmel2020}
{Brouwers} M.~G.,  {Ormel} C.~W.,  2020, \mn@doi [\aap]
  {10.1051/0004-6361/201936480}, \href
  {https://ui.adsabs.harvard.edu/abs/2020A&A...634A..15B} {634, A15}

\bibitem[\protect\citeauthoryear{{Chachan} \& {Stevenson}}{{Chachan} \&
  {Stevenson}}{2018}]{CS18}
{Chachan} Y.,  {Stevenson} D.~J.,  2018, \mn@doi [\apj]
  {10.3847/1538-4357/aaa459}, \href
  {https://ui.adsabs.harvard.edu/abs/2018ApJ...854...21C} {854, 21}

\bibitem[\protect\citeauthoryear{{Cumming}, {Helled}  \& {Venturini}}{{Cumming}
  et~al.}{2018}]{CHV18}
{Cumming} A.,  {Helled} R.,   {Venturini} J.,  2018, \mn@doi [\mnras]
  {10.1093/mnras/sty1000}, \href
  {https://ui.adsabs.harvard.edu/abs/2018MNRAS.477.4817C} {477, 4817}

\bibitem[\protect\citeauthoryear{{Dai}, {Masuda}, {Winn}  \& {Zeng}}{{Dai}
  et~al.}{2019}]{Dai19}
{Dai} F.,  {Masuda} K.,  {Winn} J.~N.,   {Zeng} L.,  2019, \mn@doi [\apj]
  {10.3847/1538-4357/ab3a3b}, \href
  {https://ui.adsabs.harvard.edu/abs/2019ApJ...883...79D} {883, 79}

\bibitem[\protect\citeauthoryear{{Dorn}, {Venturini}, {Khan}, {Heng},
  {Alibert}, {Helled}, {Rivoldini}  \& {Benz}}{{Dorn} et~al.}{2017}]{D17}
{Dorn} C.,  {Venturini} J.,  {Khan} A.,  {Heng} K.,  {Alibert} Y.,  {Helled}
  R.,  {Rivoldini} A.,   {Benz} W.,  2017, \mn@doi [\aap]
  {10.1051/0004-6361/201628708}, \href
  {https://ui.adsabs.harvard.edu/abs/2017A&A...597A..37D} {597, A37}

\bibitem[\protect\citeauthoryear{{Fegley} \& {Cameron}}{{Fegley} \&
  {Cameron}}{1987}]{FC87}
{Fegley} B.,  {Cameron} A.~G.~W.,  1987, \mn@doi [Earth and Planetary Science
  Letters] {10.1016/0012-821X(87)90196-8}, \href
  {https://ui.adsabs.harvard.edu/abs/1987E&PSL..82..207F} {82, 207}

\bibitem[\protect\citeauthoryear{{Fegley} \& {Schaefer}}{{Fegley} \&
  {Schaefer}}{2012}]{FS12}
{Fegley} Bruce J.,  {Schaefer} L.,  2012, arXiv e-prints, \href
  {https://ui.adsabs.harvard.edu/abs/2012arXiv1210.0270F} {p. arXiv:1210.0270}

\bibitem[\protect\citeauthoryear{{Freedman}, {Lustig-Yaeger}, {Fortney},
  {Lupu}, {Marley}  \& {Lodders}}{{Freedman} et~al.}{2014}]{F14}
{Freedman} R.~S.,  {Lustig-Yaeger} J.,  {Fortney} J.~J.,  {Lupu} R.~E.,
  {Marley} M.~S.,   {Lodders} K.,  2014, \mn@doi [\apjs]
  {10.1088/0067-0049/214/2/25}, \href
  {https://ui.adsabs.harvard.edu/abs/2014ApJS..214...25F} {214, 25}

\bibitem[\protect\citeauthoryear{{French}, {Becker}, {Lorenzen}, {Nettelmann},
  {Bethkenhagen}, {Wicht}  \& {Redmer}}{{French}
  et~al.}{2012}]{FrenchBecker2012}
{French} M.,  {Becker} A.,  {Lorenzen} W.,  {Nettelmann} N.,  {Bethkenhagen}
  M.,  {Wicht} J.,   {Redmer} R.,  2012, \mn@doi [\apjs]
  {10.1088/0067-0049/202/1/5}, \href
  {https://ui.adsabs.harvard.edu/abs/2012ApJS..202....5F} {202, 5}

\bibitem[\protect\citeauthoryear{{Fressin} et~al.,}{{Fressin}
  et~al.}{2013}]{F13}
{Fressin} F.,  et~al., 2013, \mn@doi [\apj] {10.1088/0004-637X/766/2/81}, \href
  {https://ui.adsabs.harvard.edu/abs/2013ApJ...766...81F} {766, 81}

\bibitem[\protect\citeauthoryear{{Fulton} et~al.,}{{Fulton}
  et~al.}{2017}]{FP17}
{Fulton} B.~J.,  et~al., 2017, \mn@doi [\aj] {10.3847/1538-3881/aa80eb}, \href
  {https://ui.adsabs.harvard.edu/abs/2017AJ....154..109F} {154, 109}

\bibitem[\protect\citeauthoryear{{Ginzburg}, {Schlichting}  \&
  {Sari}}{{Ginzburg} et~al.}{2016}]{GSS16}
{Ginzburg} S.,  {Schlichting} H.~E.,   {Sari} R.,  2016, \mn@doi [\apj]
  {10.3847/0004-637X/825/1/29}, \href
  {https://ui.adsabs.harvard.edu/abs/2016ApJ...825...29G} {825, 29}

\bibitem[\protect\citeauthoryear{{Guillot}}{{Guillot}}{1995}]{G95}
{Guillot} T.,  1995, \mn@doi [Science] {10.1126/science.7569896}, \href
  {https://ui.adsabs.harvard.edu/abs/1995Sci...269.1697G} {269, 1697}

\bibitem[\protect\citeauthoryear{{Gupta} \& {Schlichting}}{{Gupta} \&
  {Schlichting}}{2019}]{GS19}
{Gupta} A.,  {Schlichting} H.~E.,  2019, \mn@doi [\mnras]
  {10.1093/mnras/stz1230}, \href
  {https://ui.adsabs.harvard.edu/abs/2019MNRAS.487...24G} {487, 24}

\bibitem[\protect\citeauthoryear{{Harris} et~al.,}{{Harris}
  et~al.}{2020}]{numpy}
{Harris} C.~R.,  et~al., 2020, \mn@doi [\nat] {10.1038/s41586-020-2649-2},
  \href {https://ui.adsabs.harvard.edu/abs/2020arXiv200610256H} {505, 357}

\bibitem[\protect\citeauthoryear{{Hayashi}}{{Hayashi}}{1981}]{H81}
{Hayashi} C.,  1981, \mn@doi [Progress of Theoretical Physics Supplement]
  {10.1143/PTPS.70.35}, \href
  {https://ui.adsabs.harvard.edu/abs/1981PThPS..70...35H} {70, 35}

\bibitem[\protect\citeauthoryear{{Hirschmann}, {Withers}, {Ardia}  \&
  {Foley}}{{Hirschmann} et~al.}{2012}]{H12}
{Hirschmann} M.~M.,  {Withers} A.~C.,  {Ardia} P.,   {Foley} N.~T.,  2012,
  \mn@doi [Earth and Planetary Science Letters] {10.1016/j.epsl.2012.06.031},
  \href {https://ui.adsabs.harvard.edu/abs/2012E&PSL.345...38H} {345, 38}

\bibitem[\protect\citeauthoryear{{Hunter}}{{Hunter}}{2007}]{Matplotlib}
{Hunter} J.~D.,  2007, \mn@doi [Computing in Science and Engineering]
  {10.1109/MCSE.2007.55}, \href
  {https://ui.adsabs.harvard.edu/abs/2007CSE.....9...90H} {9, 90}

\bibitem[\protect\citeauthoryear{{Leconte}, {Selsis}, {Hersant}  \&
  {Guillot}}{{Leconte} et~al.}{2017}]{L17}
{Leconte} J.,  {Selsis} F.,  {Hersant} F.,   {Guillot} T.,  2017, \mn@doi
  [\aap] {10.1051/0004-6361/201629140}, \href
  {https://ui.adsabs.harvard.edu/abs/2017A&A...598A..98L} {598, A98}

\bibitem[\protect\citeauthoryear{{Ledoux}}{{Ledoux}}{1947}]{L47}
{Ledoux} P.,  1947, \mn@doi [\apj] {10.1086/144905}, \href
  {https://ui.adsabs.harvard.edu/abs/1947ApJ...105..305L} {105, 305}

\bibitem[\protect\citeauthoryear{{Lee} \& {Chiang}}{{Lee} \&
  {Chiang}}{2015}]{LC15}
{Lee} E.~J.,  {Chiang} E.,  2015, \mn@doi [\apj] {10.1088/0004-637X/811/1/41},
  \href {https://ui.adsabs.harvard.edu/abs/2015ApJ...811...41L} {811, 41}

\bibitem[\protect\citeauthoryear{{Lee}, {Chiang}  \& {Ormel}}{{Lee}
  et~al.}{2014}]{LCO14}
{Lee} E.~J.,  {Chiang} E.,   {Ormel} C.~W.,  2014, \mn@doi [\apj]
  {10.1088/0004-637X/797/2/95}, \href
  {https://ui.adsabs.harvard.edu/abs/2014ApJ...797...95L} {797, 95}

\bibitem[\protect\citeauthoryear{{Lichtenberg}, {Bower}, {Hammond},
  {Boukrouche}, {Sanan}, {Tsai}  \& {Pierrehumbert}}{{Lichtenberg}
  et~al.}{2021}]{L21}
{Lichtenberg} T.,  {Bower} D.~J.,  {Hammond} M.,  {Boukrouche} R.,  {Sanan} P.,
   {Tsai} S.-M.,   {Pierrehumbert} R.~T.,  2021, \mn@doi [Journal of
  Geophysical Research (Planets)] {10.1029/2020JE006711}, \href
  {https://ui.adsabs.harvard.edu/abs/2021JGRE..12606711L} {126, e06711}

\bibitem[\protect\citeauthoryear{{Lopez} \& {Fortney}}{{Lopez} \&
  {Fortney}}{2014}]{LF14}
{Lopez} E.~D.,  {Fortney} J.~J.,  2014, \mn@doi [\apj]
  {10.1088/0004-637X/792/1/1}, \href
  {https://ui.adsabs.harvard.edu/abs/2014ApJ...792....1L} {792, 1}

\bibitem[\protect\citeauthoryear{{Loyd}, {Shkolnik}, {Schneider},
  {Richey-Yowell}, {Barman}, {Peacock}  \& {Pagano}}{{Loyd}
  et~al.}{2020}]{LS20}
{Loyd} R.~O.~P.,  {Shkolnik} E.~L.,  {Schneider} A.~C.,  {Richey-Yowell} T.,
  {Barman} T.~S.,  {Peacock} S.,   {Pagano} I.,  2020, \mn@doi [\apj]
  {10.3847/1538-4357/ab6605}, \href
  {https://ui.adsabs.harvard.edu/abs/2020ApJ...890...23L} {890, 23}

\bibitem[\protect\citeauthoryear{{Madhusudhan}, {Nixon}, {Welbanks}, {Piette}
  \& {Booth}}{{Madhusudhan} et~al.}{2020}]{M20}
{Madhusudhan} N.,  {Nixon} M.~C.,  {Welbanks} L.,  {Piette} A. A.~A.,   {Booth}
  R.~A.,  2020, \mn@doi [\apjl] {10.3847/2041-8213/ab7229}, \href
  {https://ui.adsabs.harvard.edu/abs/2020ApJ...891L...7M} {891, L7}

\bibitem[\protect\citeauthoryear{{Markham} \& {Stevenson}}{{Markham} \&
  {Stevenson}}{2021}]{Markham21}
{Markham} S.,  {Stevenson} D.,  2021, \mn@doi [\psj] {10.3847/PSJ/ac091d},
  \href {https://ui.adsabs.harvard.edu/abs/2021PSJ.....2..146M} {2, 146}

\bibitem[\protect\citeauthoryear{{McWilliams}, {Dalton}, {Mahmood}  \&
  {Goncharov}}{{McWilliams} et~al.}{2016}]{McWilliams2016}
{McWilliams} R.~S.,  {Dalton} D.~A.,  {Mahmood} M.~F.,   {Goncharov} A.~F.,
  2016, \mn@doi [\prl] {10.1103/PhysRevLett.116.255501}, \href
  {https://ui.adsabs.harvard.edu/abs/2016PhRvL.116y5501M} {116, 255501}

\bibitem[\protect\citeauthoryear{{Misener} \& {Schlichting}}{{Misener} \&
  {Schlichting}}{2021}]{MS21}
{Misener} W.,  {Schlichting} H.~E.,  2021, \mn@doi [\mnras]
  {10.1093/mnras/stab895}, \href
  {https://ui.adsabs.harvard.edu/abs/2021MNRAS.503.5658M} {503, 5658}

\bibitem[\protect\citeauthoryear{{Mousis}, {Deleuil}, {Aguichine}, {Marcq},
  {Naar}, {Aguirre}, {Brugger}  \& {Gon{\c{c}}alves}}{{Mousis}
  et~al.}{2020}]{Mousis2020}
{Mousis} O.,  {Deleuil} M.,  {Aguichine} A.,  {Marcq} E.,  {Naar} J.,
  {Aguirre} L.~A.,  {Brugger} B.,   {Gon{\c{c}}alves} T.,  2020, \mn@doi
  [\apjl] {10.3847/2041-8213/ab9530}, \href
  {https://ui.adsabs.harvard.edu/abs/2020ApJ...896L..22M} {896, L22}

\bibitem[\protect\citeauthoryear{{Nixon} \& {Madhusudhan}}{{Nixon} \&
  {Madhusudhan}}{2021}]{NixonMadhusudhan2021}
{Nixon} M.~C.,  {Madhusudhan} N.,  2021, \mn@doi [\mnras]
  {10.1093/mnras/stab1500}, \href
  {https://ui.adsabs.harvard.edu/abs/2021MNRAS.505.3414N} {505, 3414}

\bibitem[\protect\citeauthoryear{{Ormel}, {Vazan}  \& {Brouwers}}{{Ormel}
  et~al.}{2021}]{OrmelVazan2021}
{Ormel} C.~W.,  {Vazan} A.,   {Brouwers} M.~G.,  2021, \mn@doi [\aap]
  {10.1051/0004-6361/202039706}, \href
  {https://ui.adsabs.harvard.edu/abs/2021A&A...647A.175O} {647, A175}

\bibitem[\protect\citeauthoryear{{Owen} \& {Jackson}}{{Owen} \&
  {Jackson}}{2012}]{OJ12}
{Owen} J.~E.,  {Jackson} A.~P.,  2012, \mn@doi [\mnras]
  {10.1111/j.1365-2966.2012.21481.x}, \href
  {https://ui.adsabs.harvard.edu/abs/2012MNRAS.425.2931O} {425, 2931}

\bibitem[\protect\citeauthoryear{{Owen} \& {Wu}}{{Owen} \& {Wu}}{2017}]{OW17}
{Owen} J.~E.,  {Wu} Y.,  2017, \mn@doi [\apj] {10.3847/1538-4357/aa890a}, \href
  {https://ui.adsabs.harvard.edu/abs/2017ApJ...847...29O} {847, 29}

\bibitem[\protect\citeauthoryear{{Rogers} \& {Owen}}{{Rogers} \&
  {Owen}}{2021}]{RO21}
{Rogers} J.~G.,  {Owen} J.~E.,  2021, \mn@doi [\mnras] {10.1093/mnras/stab529},
  \href {https://ui.adsabs.harvard.edu/abs/2021MNRAS.503.1526R} {503, 1526}

\bibitem[\protect\citeauthoryear{{Rogers} \& {Seager}}{{Rogers} \&
  {Seager}}{2010}]{RS10}
{Rogers} L.~A.,  {Seager} S.,  2010, \mn@doi [\apj]
  {10.1088/0004-637X/712/2/974}, \href
  {https://ui.adsabs.harvard.edu/abs/2010ApJ...712..974R} {712, 974}

\bibitem[\protect\citeauthoryear{{Rogers}, {Gupta}, {Owen}  \&
  {Schlichting}}{{Rogers} et~al.}{2021}]{RogersGupta2021}
{Rogers} J.~G.,  {Gupta} A.,  {Owen} J.~E.,   {Schlichting} H.~E.,  2021,
  \mn@doi [\mnras] {10.1093/mnras/stab2897}, \href
  {https://ui.adsabs.harvard.edu/abs/2021MNRAS.508.5886R} {508, 5886}

\bibitem[\protect\citeauthoryear{{Schaefer} \& {Fegley}}{{Schaefer} \&
  {Fegley}}{2004}]{SF04}
{Schaefer} L.,  {Fegley} B.,  2004, \mn@doi [\icarus]
  {10.1016/j.icarus.2003.08.023}, \href
  {https://ui.adsabs.harvard.edu/abs/2004Icar..169..216S} {169, 216}

\bibitem[\protect\citeauthoryear{{Schlichting} \& {Young}}{{Schlichting} \&
  {Young}}{2022}]{SY21}
{Schlichting} H.~E.,  {Young} E.~D.,  2022, \mn@doi [\psj]
  {10.3847/PSJ/ac68e6}, \href
  {https://ui.adsabs.harvard.edu/abs/2022PSJ.....3..127S} {3, 127}

\bibitem[\protect\citeauthoryear{{Scipioni}, {Stixrude}  \&
  {Desjarlais}}{{Scipioni} et~al.}{2017}]{SSD17}
{Scipioni} R.,  {Stixrude} L.,   {Desjarlais} M.~P.,  2017, \mn@doi
  [Proceedings of the National Academy of Science] {10.1073/pnas.1704762114},
  \href {https://ui.adsabs.harvard.edu/abs/2017PNAS..114.9009S} {114, 9009}

\bibitem[\protect\citeauthoryear{{Seager}, {Kuchner}, {Hier-Majumder}  \&
  {Militzer}}{{Seager} et~al.}{2007}]{S07}
{Seager} S.,  {Kuchner} M.,  {Hier-Majumder} C.~A.,   {Militzer} B.,  2007,
  \mn@doi [\apj] {10.1086/521346}, \href
  {https://ui.adsabs.harvard.edu/abs/2007ApJ...669.1279S} {669, 1279}

\bibitem[\protect\citeauthoryear{{Stevenson}, {Spohn}  \&
  {Schubert}}{{Stevenson} et~al.}{1983}]{Stevenson1983}
{Stevenson} D.~J.,  {Spohn} T.,   {Schubert} G.,  1983, \mn@doi [\icarus]
  {10.1016/0019-1035(83)90241-5}, \href
  {https://ui.adsabs.harvard.edu/abs/1983Icar...54..466S} {54, 466}

\bibitem[\protect\citeauthoryear{{Unterborn}, {Desch}, {Hinkel}  \&
  {Lorenzo}}{{Unterborn} et~al.}{2018}]{U18}
{Unterborn} C.~T.,  {Desch} S.~J.,  {Hinkel} N.~R.,   {Lorenzo} A.,  2018,
  \mn@doi [Nature Astronomy] {10.1038/s41550-018-0411-6}, \href
  {https://ui.adsabs.harvard.edu/abs/2018NatAs...2..297U} {2, 297}

\bibitem[\protect\citeauthoryear{{Valencia}, {O'Connell}  \&
  {Sasselov}}{{Valencia} et~al.}{2006}]{V06}
{Valencia} D.,  {O'Connell} R.~J.,   {Sasselov} D.,  2006, \mn@doi [\icarus]
  {10.1016/j.icarus.2005.11.021}, \href
  {https://ui.adsabs.harvard.edu/abs/2006Icar..181..545V} {181, 545}

\bibitem[\protect\citeauthoryear{{Vazan} \& {Helled}}{{Vazan} \&
  {Helled}}{2020}]{Vazan2020}
{Vazan} A.,  {Helled} R.,  2020, \mn@doi [\aap] {10.1051/0004-6361/201936588},
  \href {https://ui.adsabs.harvard.edu/abs/2020A&A...633A..50V} {633, A50}

\bibitem[\protect\citeauthoryear{{Vazan}, {Helled}  \& {Guillot}}{{Vazan}
  et~al.}{2018}]{VHG18}
{Vazan} A.,  {Helled} R.,   {Guillot} T.,  2018, \mn@doi [\aap]
  {10.1051/0004-6361/201732522}, \href
  {https://ui.adsabs.harvard.edu/abs/2018A&A...610L..14V} {610, L14}

\bibitem[\protect\citeauthoryear{{Vazan}, {Sari}  \& {Kessel}}{{Vazan}
  et~al.}{2022}]{VazanSari22}
{Vazan} A.,  {Sari} R.,   {Kessel} R.,  2022, \mn@doi [\apj]
  {10.3847/1538-4357/ac458c}, \href
  {https://ui.adsabs.harvard.edu/abs/2022ApJ...926..150V} {926, 150}

\bibitem[\protect\citeauthoryear{{Virtanen} et~al.,}{{Virtanen}
  et~al.}{2020}]{scipy}
{Virtanen} P.,  et~al., 2020, \mn@doi [Nature Methods]
  {10.1038/s41592-019-0686-2}, \href
  {https://ui.adsabs.harvard.edu/abs/2020NatMe..17..261V} {17, 261}

\bibitem[\protect\citeauthoryear{{Visscher} \& {Fegley}}{{Visscher} \&
  {Fegley}}{2013}]{VF13}
{Visscher} C.,  {Fegley} Bruce J.,  2013, \mn@doi [\apjl]
  {10.1088/2041-8205/767/1/L12}, \href
  {https://ui.adsabs.harvard.edu/abs/2013ApJ...767L..12V} {767, L12}

\bibitem[\protect\citeauthoryear{{Wahl} et~al.,}{{Wahl} et~al.}{2017}]{W17Juno}
{Wahl} S.~M.,  et~al., 2017, \mn@doi [\grl] {10.1002/2017GL073160}, \href
  {https://ui.adsabs.harvard.edu/abs/2017GeoRL..44.4649W} {44, 4649}

\bibitem[\protect\citeauthoryear{{Weiss} \& {Marcy}}{{Weiss} \&
  {Marcy}}{2014}]{WM14}
{Weiss} L.~M.,  {Marcy} G.~W.,  2014, \mn@doi [\apjl]
  {10.1088/2041-8205/783/1/L6}, \href
  {https://ui.adsabs.harvard.edu/abs/2014ApJ...783L...6W} {783, L6}

\bibitem[\protect\citeauthoryear{{Zeng} et~al.,}{{Zeng}
  et~al.}{2019}]{Zeng2019}
{Zeng} L.,  et~al., 2019, \mn@doi [Proceedings of the National Academy of
  Science] {10.1073/pnas.1812905116}, \href
  {https://ui.adsabs.harvard.edu/abs/2019PNAS..116.9723Z} {116, 9723}

\makeatother
\end{thebibliography}

% Alternatively you could enter them by hand, like this:
% This method is tedious and prone to error if you have lots of references
%\begin{thebibliography}{99}
%\bibitem[\protect\citeauthoryear{Author}{2012}]{Author2012}
%Author A.~N., 2013, Journal of Improbable Astronomy, 1, 1
%\bibitem[\protect\citeauthoryear{Others}{2013}]{Others2013}
%Others S., 2012, Journal of Interesting Stuff, 17, 198
%\end{thebibliography}

%%%%%%%%%%%%%%%%%%%%%%%%%%%%%%%%%%%%%%%%%%%%%%%%%%

%%%%%%%%%%%%%%%%% APPENDICES %%%%%%%%%%%%%%%%%%%%%

%%%%%%%%%%%%%%%%%%%%%%%%%%%%%%%%%%%%%%%%%%%%%%%%%%

% Don't change these lines
\bsp	% typesetting comment
\label{lastpage}
\end{document}